\documentclass[12pt,preprint]{aastex}
\usepackage{graphicx}
\usepackage{epstopdf}
\def\ls{{_<\atop^{\sim}}}
\def\gs{{_>\atop^{\sim}}}

\def\um{~$\mu$m}

\def\h0{H$_0$}
\def\q0{q$_0$}

\def\Msun{M$_{\odot}$}
\def\Lsun{L$_{\odot}$}

\def\han {\mbox{{\rm H}$\alpha$}}

\def\hii{\mbox{H~{\sc ii}}}
\def\ha{\han}

\def\brg {\mbox{{\rm Br}$\gamma$}}

\def\kms {\mbox{{\rm km} {\rm s}$^{-1}$}}
\def\spose#1{\hbox to 0pt{#1\hss}}
\def\simlt{\mathrel{\spose{\lower 3pt\hbox{$\mathchar''218$}}
     \raise 2.0pt\hbox{$\mathchar''13C$}}}
\def\simgt{\mathrel{\spose{\lower 3pt\hbox{$\mathchar''218$}}
     \raise 2.0pt\hbox{$\mathchar''13E$}}}
\def\lsim{\rlap{$<$}{\lower 1.0ex\hbox{$\sim$}}}
\def\gsim{\rlap{$>$}{\lower 1.0ex\hbox{$\sim$}}}
\def\be{\begin{equation}}
\def\ee{\end{equation}}

\slugcomment{\it Accepted to ApJ}

\shortauthors{Gilbert \& Graham}
\shorttitle{Feedback from Antennae ELCs I.}

\begin{document}

\title{Feedback in the Antennae Galaxies (NGC 4038/9): 
I. High-Resolution Infrared Spectroscopy of Winds from Super Star
Clusters\altaffilmark{1,2}}

\author{Andrea M. Gilbert\altaffilmark{3,4,5} \&
James R. Graham\altaffilmark{4}
}  

\altaffiltext{1}{Data presented herein were obtained
at the W.M. Keck Observatory, which is operated as a scientific
partnership among the California Institute of Technology, the
University of California and the National Aeronautics and Space
Administration.  The Observatory was made possible by the generous
financial support of the W.M. Keck Foundation.}

\altaffiltext{2}{Data presented herein were obtained at the European 
Southern Observatory Very Large Telescope.}

\altaffiltext{3}{Institute of Geophysics \& Planetary Physics, Lawrence Livermore National Laboratory, L-413, 7000 East Ave., Livermore, CA 94550}

\altaffiltext{4}{Department of Astronomy,  
University of California,
601 Campbell Hall,
Berkeley, CA, 94720-3411, USA}

%\altaffiltext{4}{Max-Planck Institut f\"ur extraterrestrische Physik,
%Postfach 1312, D-85741 Garching, Germany}

%\altaffiltext{5}{agilbert@mpe.mpg.de}

\altaffiltext{5}{agilbert@igpp.ucllnl.org}

%%%%%%%%%%%%
% Abstract %
%%%%%%%%%%%%

\begin{abstract}
 We present high-resolution ($R \sim 24,600$) near-IR spectroscopy of
 the youngest super star clusters (SSCs) in the prototypical starburst
 merger, the Antennae Galaxies.  These SSCs are young ($3-7$ Myr old)
 and massive ($10^5 - 10^7$ M$_\odot$ for a Kroupa IMF) and their
 spectra are characterized by broad, extended Brackett$\gamma$
 emission, so we refer to them as emission-line clusters (ELCs) to
 distinguish them from older SSCs.  The Br$\gamma$ lines of most ELCs
 have supersonic widths ($60-110$ km s$^{-1}$ FWHM) and non-Gaussian
 wings whose velocities exceed the clusters' escape velocities.  This
 high-velocity unbound gas is flowing out in winds that are powered by
 the clusters' massive O and W-R stars over the course of at least
 several crossing times.  The large sizes of some ELCs relative to
 those of older SSCs may be due to expansion caused by these outflows;
 many of the ELCs may not survive as bound stellar systems, but rather
 dissipate rapidly into the field population.  The observed tendency
 of older ELCs to be more compact than young ones is consistent with
 the preferential survival of the most concentrated clusters at a
 given age.
 \end{abstract} 


%%%%%%%%
% Body %
%%%%%%%%

\section{Introduction}
\label{intro}

Star formation in starbursts creates massive young super star clusters
(SSCs) that are not often found in more quiescent environments like
the Milky Way's disk.  Their inferred masses ($10^5-10^6$ \Msun) and
Lyman continuum photon rates ($10^{52}-10^{53}$ s$^{-1}$) dwarf those of their
lower-mass analogs in the Local Group.  The most massive local SSC
analog is R136 in 30~Doradus, with a mass of at least
$10^{4.5}$ \Msun\  \citep{massey98a} and  $10^{51.4}$ s$^{-1}$ \citep{walborn91}.

SSCs are found in all galaxy environments, from dwarf irregular
starbursts \citep[e.g. NGC 1569,][]{oconnell94} to the prototypical
starburst disk galaxy M82 \citep[e.g.][]{oconnell95}, to
merging systems such as the Antennae Galaxies (NGC~4038/39)
\citep[e.g.][]{whitmore95,gilbert02}, and some have been found in
normal spiral galaxies  \citep[like NGC~6946][]{larsen01}, whose young clusters typically populate a lower mass range, $\sim 10^3 - 10^5$ \Msun\  \citep[e.g.][]{larsen02}.
% added citation/comment
Most
observations of SSCs have focused on optical wavelengths, but
near-infrared (near-IR) imaging has revealed young massive clusters
that excite luminous H~{\sc ii} regions (HIIRs) that are
invisible at optical wavelengths \citep[e.g.][]{gilbert00,turner03}.
Narrow-band imaging may also be important to track the faintest
youngest members of the population, which may not be detected with
broad-band IR imaging although they are bright in narrow filters tuned
to the recombination lines of their HIIRs \citep{alonso02}.  Radio
and mid-IR observations have uncovered still more heavily embedded
regions of massive star formation that are heavily obscured even at
2 $\mu$m \citep[e.g.][]{kobulnicky99,beck00,beck01,vacca02,cabanac05}; these objects
are very young, dense, compact HIIRs that are the massive analogs of
ultra-compact \hii\ regions (UCHIIs).  Hence they are dubbed
ultra-dense HIIRs, or UDHIIs, by \citet[][]{kobulnicky99}, and they are
the supposed precursors of optically visible SSCs.

The high masses and stellar densities of SSCs resemble those of
globular clusters (GCs), although metal abundances ought to be higher in SSCs than GCs
and SSC ages are measured in Myr rather
than Gyr.  This suggests an evolutionary sequence in massive
star-forming regions: UDHIIs become optically visible SSCs
\citep{kobulnicky99} that may finally become (metal-rich) GCs.  The idea of GC
formation in mergers is supported by observations of young GCs (ages
of a few Gyr) in several merger remnants
\citep[e.g.][]{schweizer87,ashman92,fritzevalvensleben94,schweizer96,schweizer98},
as well as the multi-modal color distributions and varied spatial
distributions of GCs.  However, establishing the relationship between
SSCs and GCs requires examination of both the photometric and
kinematic properties of the SSC population.  By measuring an
individual SSC's age, mass, density profile and initial mass function
(IMF), and modeling its stellar and dynamical evolution, one can
predict whether it may survive internal and galactic dynamical
processes to join a population of GCs
\citep[e.g.][]{ho96a,boeker99,smith01,larsen01,mengel02,gilbert02,mccrady03}.
The IMF of SSCs is especially important for this question because the
light of SSCs is dominated by massive stars, while old GCs are
comprised entirely of sub-solar mass stars (and stellar remnants).  Observations of SSC
dynamical masses to date suggest that some SSCs do have normal IMFs
and could become GCs, while others cannot
\citep[e.g.][]{sternberg98,smith01,mengel02,gilbert01b,mccrady03,mccrady05},
although these results could be subject to systematic errors in
extinction corrections or neglect of effects like spatial mass
segregation, and they always rely on the assumption of cluster
virialisation.

Another important issue is the evolution of the SSC population as whole: GC
populations have kinematic distributions resembling those of bulges
(spheroidal with high velocity dispersions), while SSCs form where gas is
available, usually in disk-like environments with much lower dispersions; GCs
have log-normal luminosity functions \citep{harris91}, while SSCs display
power-law luminosity functions \citep[e.g.][]{whitmore99}, as do the giant
molecular clouds (GMCs) and \hii\ regions from which they presumably form
\citep[e.g.][]{williams97,kennicutt88,mckee97}.  Simulations have shown that
power-law cluster LFs can evolve into log-normal ones as the faint end of the
LF is eroded by tidal disruption in the galactic
potential, dynamical friction, stellar mass loss, and evaporation of stars
from clusters \citep[e.g.][]{fall01}.

Whether young SSCs survive to become GCs or disperse completely into
the field star population of a galaxy, they have great potential to
affect the energetics of its interstellar medium (ISM) because they
harbor thousands of massive stars producing ionizing and FUV
radiation.  The radiation from OB stars excites HIIRs and
photodissociation regions (PDRs), and heats the ISM.  Their winds and
supernova (SN) ejecta stir and inject energy into the surrounding ISM.
In the most extreme starbursts, the combined effects of a starburst
drive large-scale galactic winds \citep[e.g. in M82,][]{shopbell98}
that can eject matter into the intergalactic medium \citep[especially
in dwarf galaxies, e.g. NGC 1569,][]{heckman95}. SSCs are concentrated
power sources for feedback on both star-cluster and galactic scales.

The Antennae (NGC~4038/9) are a nearby pair of disk galaxies in an
early stage of merging that are well-known for their numerous SSCs,
distributed along their spiral arms and around their interaction
region \citep{whitmore99}.  Their molecular gas
distribution peaks at both nuclei and in the overlap region
\citep{stanford90}, but the gas is not yet undergoing a global
starburst typical of more advanced mergers \citep{nikola98}.  The
current global star-formation rate of the system is estimated at about
20 \Msun\ yr$^{-1}$ by \citet{zhang01} from extinction-corrected
H$\alpha$.  ISO observations of the Antennae show that the mid-IR
($8-15$ \um) flux from warm dust follows the optical distribution of
bright blue SSCs, but peaks in the overlap region in a single point
source that emits 15\% of the ISO flux \citep{mirabel98}.  This source
is an optically faint and unremarkable star cluster \citep{whitmore95}
that near-IR observations reveal as one of the youngest ($\sim 4$
Myr), most massive ($\sim 10^7$ \Msun) SSCs in the system
\citep{gilbert00,mengel01a}.  This cluster powers a dense (average
$n_e= 10^4$), large (half-light radius 32 pc) \hii\ region with a
Lyman continuum rate of order $Q[H^+]=10^{53}$ s$^{-1}$; it excites even
more extended, clumpy PDRs with nearly fluorescent H$_2$ emission; and
it is embedded behind $A_V \approx 10$ mag \citep{gilbert00}.

In \S~\ref{sec:obselc} we present near-IR imaging and spectroscopic
observations of a sample of young Antennae SSCs in order to identify
the emission-line clusters (ELCs) among them.  We derive ages and
stellar masses for the ELCs from their \brg\ spectra and magnitudes in
\S~\ref{sec:agemass}.  In \S~\ref{sec:linewidths} we compare the
widths of the broad, extended \brg\ emission lines of ELCs with the
inferred escape velocities of the exciting SSCs, and argue that ELCs
drive cluster outflows.  In \S~\ref{sec:densities} we estimate
electron densities for ELCs, and we compare ELCs with other types of
HIIR.  We discuss the evolution of ELCs and survival of massive clusters undergoing expansion due to mass loss in \S~\ref{sec:evolution}, estimate the ELC contribution to the
total star-formation rate in the Antennae in \S~\ref{sec:sfr}, and
conclude in \S~\ref{sec:conclusions}.  In a companion paper 
\citep[Paper II,][]{gilbert07b}, we present
a kinematic model for the cluster outflows, fit it to the data presented here,
 and discuss the mass-loss rates and energy budget of ELC outflows
 relative to similar systems such as superbubbles and galactic superwinds.

\section{Observations \& Data Reduction}
\label{sec:obselc}

On 2002 January 21 and February 22, we obtained $K$-band spectra of a
sample of 17 young SSCs and the nuclei of NGC~4038 and NGC~4039 in the
Antennae using NIRSPEC \citep{mclean00}, a facility near-infrared
($0.95 - 5.6 \mu$m) spectrometer for the Keck-II telescope
\citep{mclean98}.  Using the cross-dispersed echelle mode and the N7
order-sorting filter, we detected Brackett$\gamma$ with little or no
continuum in 16 of the targets with $\lambda/\Delta\lambda \simeq 24,600$.  During the
course of these observations we recorded N7 images of the fields
around the $0.''432$ (3-pixel) slit using NIRSPEC's slit-viewing
camera (SCAM); these images were used to create a mosaicked finding
chart of the Antennae field from which offsets between the infrared
clusters were measured.  Cluster photometry was performed on $K$-band
images taken with NIRSPEC's slit-viewing camera (SCAM) in good seeing
($0.''6$ or better) and photometric conditions on 2000 January 17 and
2001 March 10.  We later obtained higher quality VLT ISAAC
\citep{moorwood98} narrow-band images at 2.17 and 2.25 \um\ on 11 January 2004 
in K-band seeing of $0\farcs4-0\farcs5$;
the former filter includes \brg\ emission and is shown in the finding chart of Figure~\ref{fig:ant_hst}.
% added ISAAC seeing here

\subsection{Imaging}
\label{sec:irlf}

\begin{figure}[h] 
\begin{center}
%\epsscale{0.85} 
\plotone{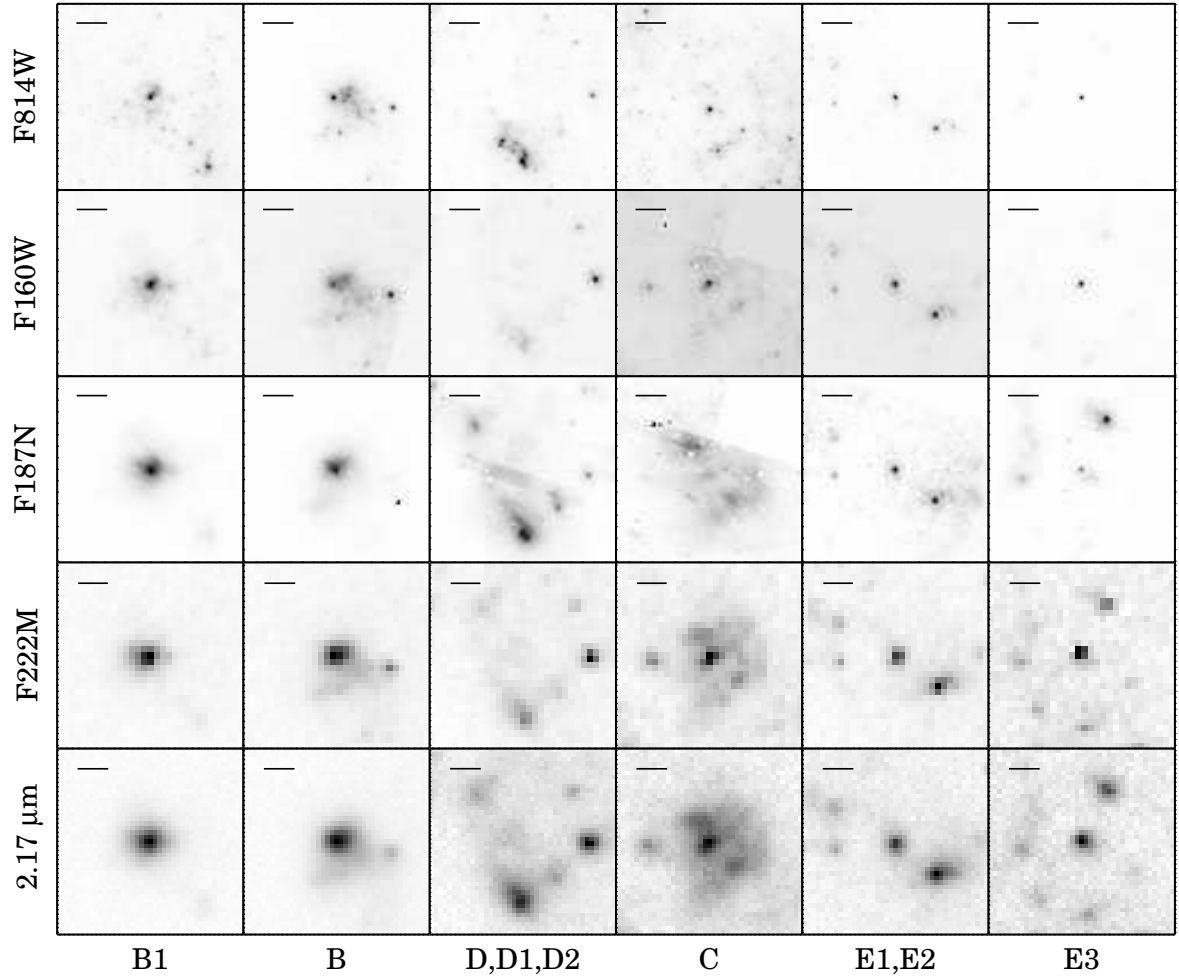}
\caption[Comparison of ELC images from ISAAC and HST] {Comparison of ELCs observed from the ground and by HST in several wavebands.  Bottom row shows the ISAAC 2.17 \um\ image (\protect{0\farcs4} seeing); subsequent rows show NIC3 F222M image (\protect{0\farcs2/}pixel), NIC2 F187N and F160W images (\protect{0\farcs075}/pixel), and ACS F814W image (\protect{0\farcs05}/pixel).  Horizontal bar is \protect{1\arcsec}\ long. 
\label{fig:thumbs} }
\end{center}
\end{figure}

The bright northern and southern sources in Figure~\ref{fig:ant_hst}
are the two nuclei in the system, NGC~4038 and NGC~4039.  Overplotted
are the positions of the brightest young ($< 30$ Myr old) clusters
detected in optical HST images reported by \citet{whitmore99}.  The
HST and NIRSPEC data sets were registered using the foreground star at
the origin of the figure (Star 4 of Whitmore et al.
1999)\footnote{\citet{whitmore99} report a position for Star 4 of RA =
  $12^h 1^m 56.04^s$ and DEC = $-18^\circ 52' 43\farcs66$ (J2000),
  although an offset of $1\farcs2$ to the southwest of the HST positions
  from radio observations \cite{neff00} was reported by
  \citet{whitmore02}.}.  Comparing the relative HST positions of the
nine cleanly overlapping sources with their relative IR positions
yielded a SCAM plate scale of $0\farcs178$ per pixel.  This is
consistent with the value of $0\farcs18$ measured by \citet{figer00}. 
% modify opt/ir coincidence...
%The coincidence of so few bright optical sources in Figure~\ref{fig:ant_hst} with bright infrared ones reveals the incompleteness of the optical observations due to dust extinction in the region between the nuclei.  This highlights the importance of long-wavelength observations in probing the most recent star formation. 
About half of our sample of bright young SSCs does not appear in the optical brightest
   cluster list due to dust extinction, although most of these sources are
   detected at least in F814W (see Figure~\ref{fig:thumbs}).   SSC B1 highlights the importance of
   long-wavelength observations in probing the most recent star
   formation: it is deeply embedded and quite faint in the optical,
   with a K-band extinction of at least 1 mag \citep{gilbert00},
   and it is the most massive young SSC in the system after correcting
   for this extinction (Table~\ref{tab:agemass}).

The $K$-band SCAM images were reduced and calibrated relative to the
\citet{persson98} standards SJ 9150 and SJ 1952.  Aperture magnitudes for the
clusters were calculated using DAOPhot \citep{stetson87} routines in IDL.  The
fields containing the clusters are complex and it is evident that the accuracy
of our photometry is limited by systematics associated with eliminating
contributions from unrelated sources and with estimating the local background.
In an attempt to reduce these systematic errors we also computed magnitudes
based on fitting a 2-d Gaussian to the resolved cluster light profile.  The
mean difference between the $2''$ diameter aperture and fitting magnitudes is
$0.12$ mag with an rms of 0.2 mag.  Inspection of the field confirms that the
$2''$ apertures represent a compromise between isolating the clusters from
their environment while encompassing most of the light  \citep[point-source aperture
corrections were applied using Star 4 of][]{whitmore99}. 
We therefore adopt these magnitudes in what follows
and assume 0.2 mag as the rms error.  Table~\ref{tab:phottable} lists the
magnitudes of the selected targets for spectroscopy.  Our photometry agrees
with other measurements from the literature: for SSC B1 \citep[known as SSC A by][]{gilbert00}, we find K = 14.7 $\pm$
0.2 mag, \citet{gilbert00} report N7 = 14.6 mag, and \citet{mengel01a} report
K = $14.8 \pm 0.2$ mag; for SSC F we find K $=15.9 \pm$ 0.2 mag
 and \citet{mengel02} report K$_s = 15.9$ mag.  For SSC S 
 we did not have adequate data for photometry, so we use values reported by
 \citet{mengel02}.  
 
 % added detail about isaac sizes here. 
 The ISAAC images were reduced, corrected for distortion, and combined.  Sizes of ELCs and Star 4 were measured via 2D Gaussian and Lorentzian fits, which gave comparable results, and the latter were adopted because they better reproduced the observed profiles.  We derived FWHMs from the geometric mean radii of the fits because it is a better indicator than the arithmetic mean of the area in an image that encloses half of the profile's light.  The FWHM of Star 4 was subtracted in quadrature to
compute the deconvolved FWHMs that are reported in Table~\ref{tab:phottable}
and plotted in Figure~\ref{fig:sizes}.
 Throughout this paper we adopt a redshift distance for the Antennae of 19.2 Mpc ($1\arcsec = 93$ pc, distance modulus $(m-M)_0=31.41$ mag for $H_0=75$ km s$^{-1}$ Mpc$^{-1}$), although we note that a smaller distance of 13.8 Mpc has been measured by \citet{saviane04} based on the tip of the red giant branch (TRGB) in the southern tail of the Antennae.  If the TRGB distance is correct, all linear dimensions discussed in this paper decrease by a factor of 1.4.

\subsection{Brackett$\gamma$ Spectroscopy}
\label{sec:brg}

\begin{figure}[ph]
\begin{center}
\plotone{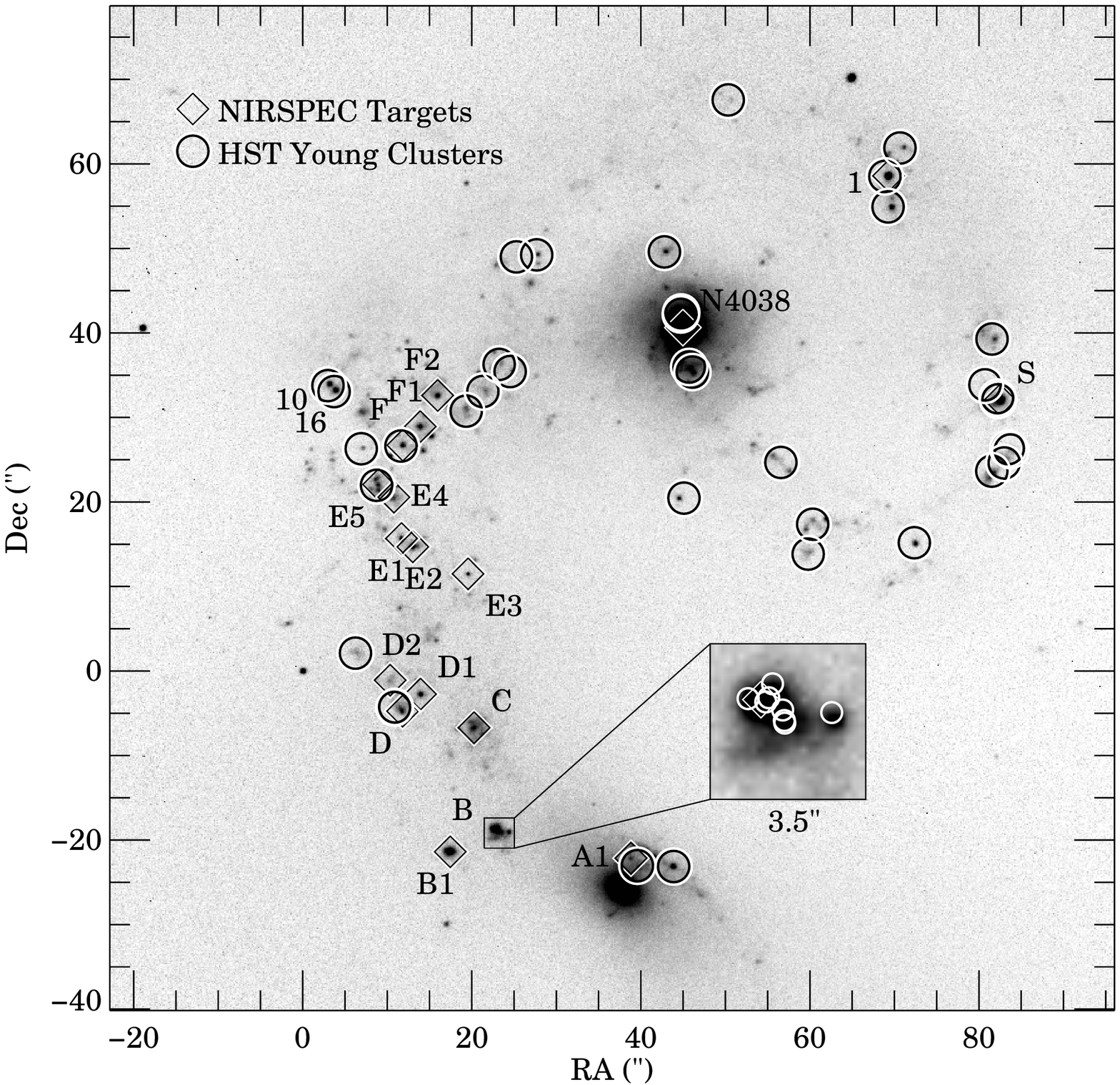}
\caption[ISAAC 2.17 \um\ image of the Antennae] {Image of the Antennae using
  ISAAC 2.17 \um\ narrow-band filter, which includes \brg.  Units are
  arcseconds of R.A. and Dec.; north is up and east is to the left.  Diamonds mark positions of NIRSPEC targets (except for the circled sources 10 and 16 where they are omitted to avoid crowding).  Circles mark positions of the brightest young clusters in HST F555W (Whitmore et al.
  1999).  SSC B is shown in an inset, magnified by a factor of five, in order to reveal IR structure and separate HST sources.  At the origin is Star 4 of \citet{whitmore99}. 
\label{fig:ant_hst}}
\end{center}
\end{figure}

A finding chart similar to that in Figure~\ref{fig:ant_hst} from
prior $K$-band observations permitted the selection of cluster targets
for spectroscopy. Targets were chosen on the basis of brightness;
whenever feasible we oriented the $24''$ long slit to encompass
multiple clusters.

Table~\ref{tab:obselc} lists exposure times for the targets that are
identified in the image of the Antennae in Figure~\ref{fig:ant_hst}.  
SSCs are named roughly according to the lettering scheme adopted for \ha-emitting regions in the system by \citet{rubin70}.
Along with the SSCs we observed the two nuclei and detected broad
emission in the northern nucleus, NGC 4038, but not in the southern
one, NGC 4039, although we did detect broad Br$\gamma$ in a faint cluster near
the latter nucleus.  This off-nuclear source is labeled ``A1'' in
Figure~\ref{fig:ant_hst} and in Table~\ref{tab:measurements}.

Raw exposures were first corrected for electronic noise which appears
as horizontal striping in a few of the 32 analog channels.  The images
were then dark-subtracted (because on-minus-off sky subtraction was not
possible for a few exposures),
 flat-fielded, and corrected for bad pixels
and cosmic rays.  A wavelength scale was determined from atmospheric
OH lines \citep{rousselot00} and used to rectify the Brackett $\gamma$ echelle order
(order 35), in which spatial and spectral directions are curved, onto
an orthogonal grid of wavelength and slit position.  In order 35 the
spectral resolution measured from OH lines was
${\lambda}/{\delta\lambda}= 24,600$, or about 12 km s$^{-1}$ FWHM.

A primary sky subtraction was performed by differencing pairs of
frames where possible, and then a secondary sky subtraction was done
by fitting and removing the background in each column.
Emission-line spectra were optimally extracted using a Gaussian
weighting function matched to the spatial extent of the Br$\gamma$
emission, and then an aperture correction was applied to recover the
full flux.

\begin{figure}[ph] 
\begin{center}
\epsscale{0.85} 
% was .65
\vbox{
\plottwo{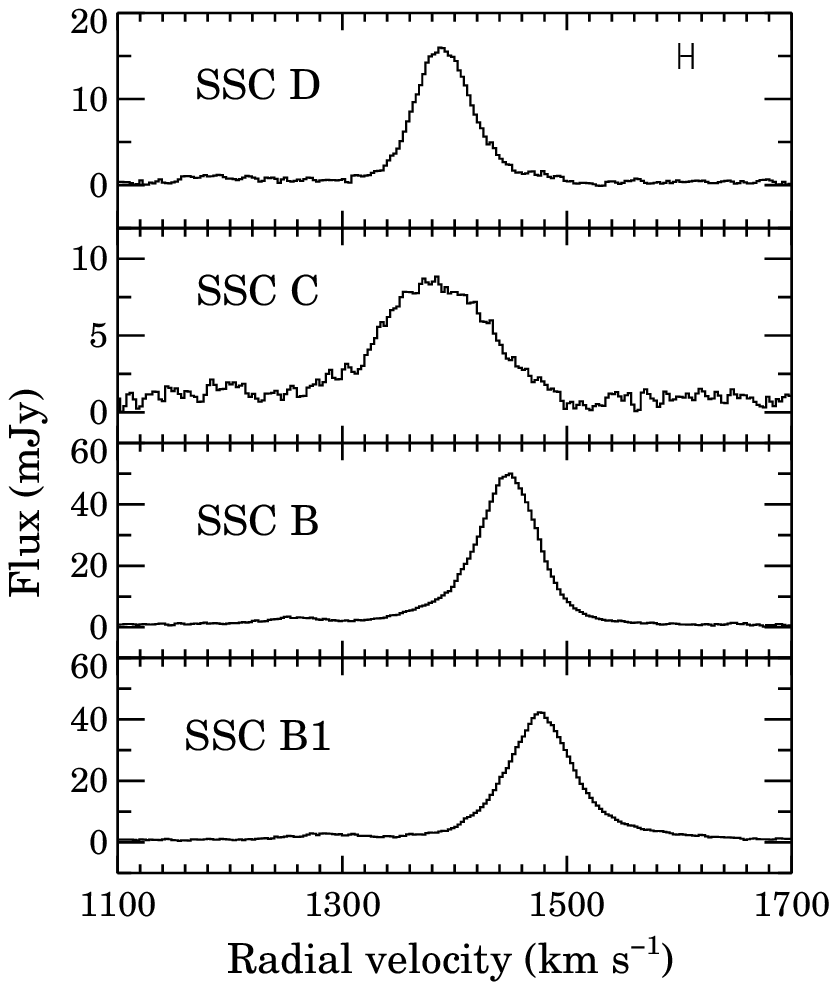}{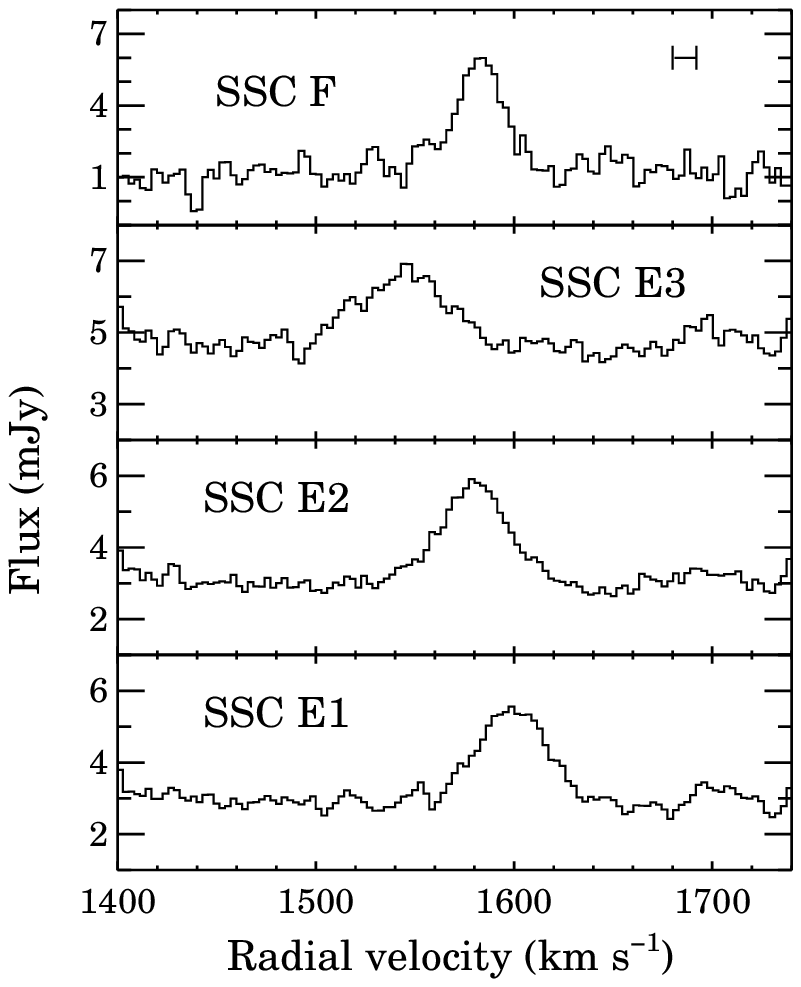}
% was 1.0
\epsscale{0.85}
\plottwo{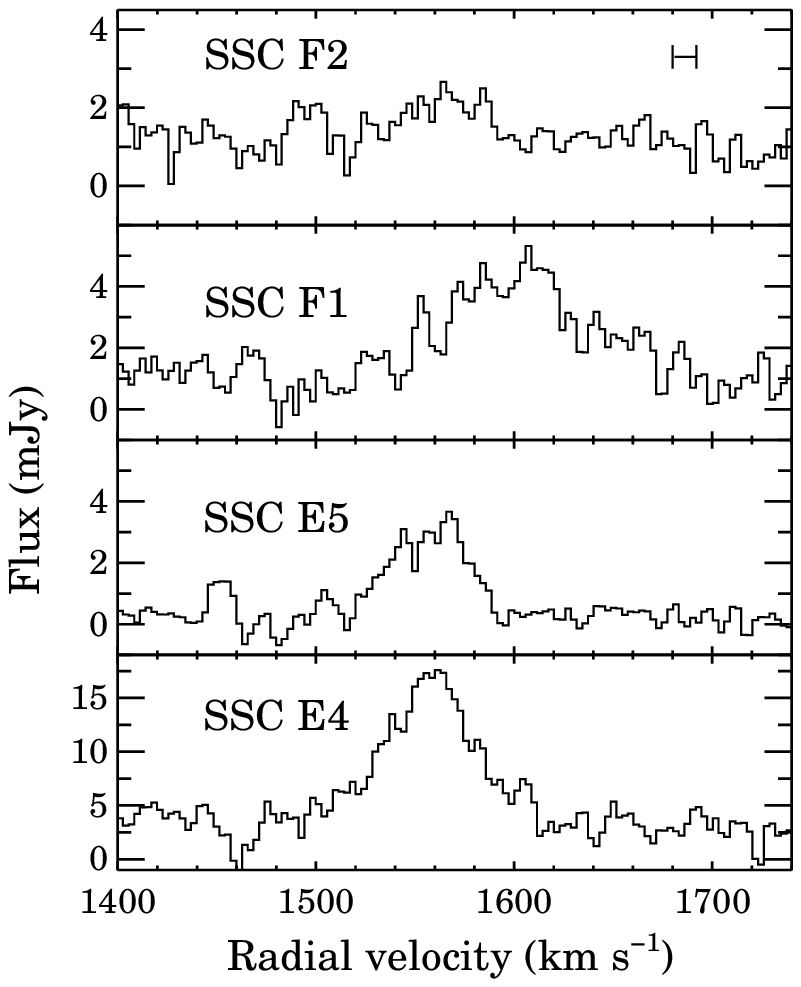}{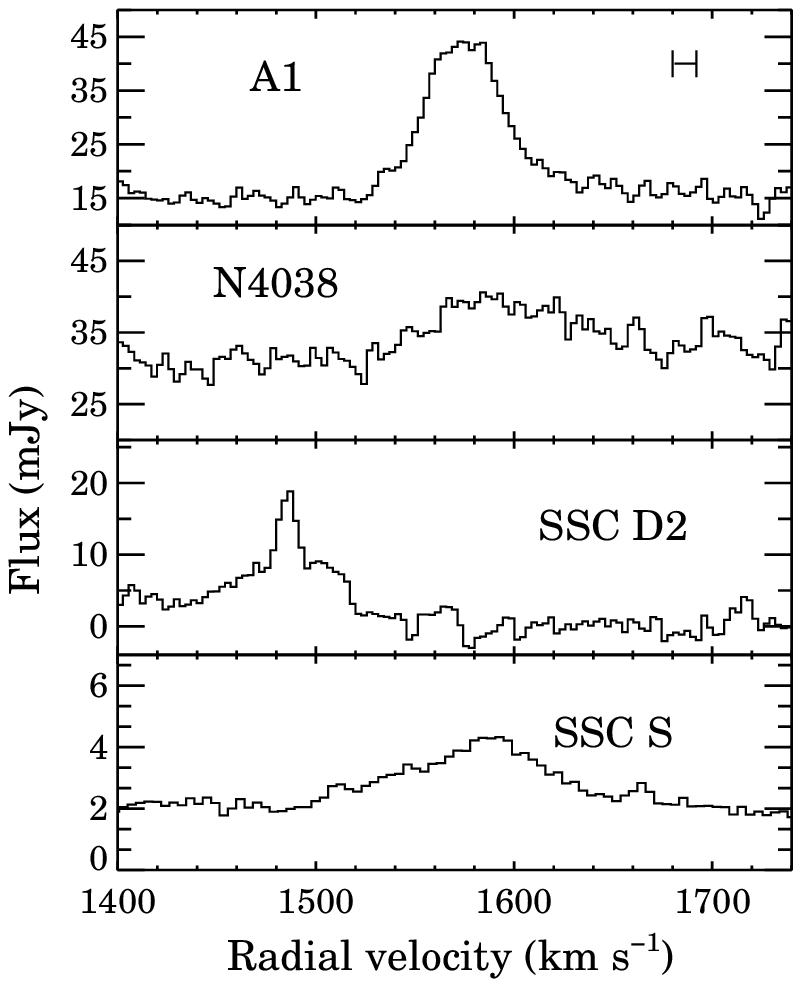}
}
\caption[Br$\gamma$ echelle spectra of Antennae targets] {Br$\gamma$ spectra
  of young Antennae SSCs and nuclear regions, featuring broad,
  non-Gaussian wings.  The spectral resolution is about 12 \kms\ FWHM (4
  pixels, shown as a bar in upper right corner of all panels), so the lines are well resolved and their widths are not
  simply due to thermal broadening.  He~{\sc i} emission at 2.1647
  $\mu$m ($\Delta v = -191$ \kms from \brg) is also visible in the
  brightest sources (top left panel). Note the different velocity
  scale in top left panel, and varied flux scales throughout.}
\label{fig:brg}
\end{center}
\end{figure}

A B4IV/V star was observed as an atmospheric standard.  Although most
of our Br$\gamma$ spectra were unaffected by atmospheric absorption,
at the radial velocities of some objects a small feature absorbs some
flux in the wing of the line, so we applied the atmospheric correction
in all spectra.  A flux scale was derived by requiring the 2.2 $\mu$m
continuum flux of the standard star to equal that corresponding to its K
magnitude.  Reduced line spectra for the ELCs from which we
detected broad Br$\gamma$ (excepting only D1) are shown in Figure~\ref{fig:brg}, and
position-velocity diagrams for the four brightest ELCs are shown in
Figure~\ref{fig:elcpv}.  Signal-to-noise ratios per pixel in the fully
reduced spectra ranged from around 3.5 in the weakest lines to 75 in
the strongest, with continuum levels  low in all but the brightest
sources, which have continuum signal-to-noise ratios per pixel of only a few.
Sources for which we did not detect \brg, clusters 10, 16, and 1 from 
\citet{whitmore99}, are not considered further in this paper.
We refer to the SSCs with \brg\ emission as emission-line clusters (ELCs).

The Br$\gamma$ profiles in Figure~\ref{fig:brg} are well resolved and
exhibit high-velocity non-Gaussian wings.  The isothermal sound speed
of ionized gas at temperature $T=10^4$ K is $ c =\sqrt{k_B T/\mu m_p}
\approx $ 12 km s$^{-1}$ where $\mu$ is the mean molecular weight and
$m_p$ is the proton mass, so linewidths FWHM = $2.355 c > 28 $ \kms\ 
are supersonic. Nearly all of the line widths, determined from
Gaussian fits to the line cores, are supersonic. Cluster D1 (FWHM = 14
km s$^{-1}$) is the only Br$\gamma$ source without a supersonic
component and cluster F is marginal with FWHM = $30$ \kms.  We refer
to these young SSCs with supersonic Br$\gamma$ line widths as broad
emission-line clusters (bELCs). 

The broad Br$\gamma$ lines of ELCs may arise from individual hot stars
or dilute photoionized gas.  Individual O stars have EW[\brg] of only
a few \AA~\citep[e.g.][]{hanson96}, while the Antennae clusters have
EW[\brg] from a few up to a few hundred \AA.  The observed Br$\gamma$
lines are also too narrow to be associated with individual O-star
winds, and as already noted, they are too broad (and non-Gaussian) to
simply reflect the virial motion of the constituent stars.  Thus,
photospheric and wind emission from O stars alone cannot explain the
intensity or width of \brg\ emission from ELCs.  Even though the ELC
lines are too narrow to be stellar wind features, WR stars could
produce the observed equivalent widths.  \citet{figer97} measure the
combined EW of blended He~{\sc i}, He~{\sc ii} and \brg\ lines from
low-resolution IR spectra of WR stars, finding values of about 20--130
\AA.  However, the line widths are far too narrow to be dominated by
WR emission \citep[which has line widths $\sim 10^3$ \kms,
e.g.][]{abbott87}.  Thus the ELC line fluxes most likely originate not from gas in photospheres
and winds of hot stars, but in the expanding \hii\ regions that they excite.
The broad lines must be formed from rapidly expanding ionized gas swept up by
the cumulative effect of stellar winds and photoevaporated flows driven by the
thousands of hot stars in the clusters.

% add comparison to Whitmore05 Ha radial velocities
Table~\ref{tab:measurements} summarizes the measured Br$\gamma$
fluxes, equivalent widths, barycentric radial velocities, and line
widths for the ELCs.  
Our \brg\ radial velocities are consistent with those reported from HST \ha\ spectroscopy by \citet{whitmore05} for sources in the vicinities of SSCs B, D, and F (a direct cluster-to-cluster comparison is not possible because the slit positions and sizes are different).
The listed \brg\ FWHMs are the observed values, not corrected
for either the instrumental line-spread function (12 \kms) or
thermal broadening (21.4 \kms).
We also detected He~{\sc i} 3F$^0-$3G emission at 2.1647 $\mu$m near
the blue wing of Br$\gamma$ in the brightest sources (B1, B, C, and D)
at a level of $4.4 \pm 0.5$\% of the \brg\ fluxes.  For SSC D2, whose
Br$\gamma$ emission includes a very narrow component (FWHM = 23 \kms)
as well as a weak broad component (FWHM = 55 \kms), we list the total
line flux but only the line width of the broad part.

% change gerssen reference to bottema; add more dispersion discussion
We consider the kinematics of the ELC population in the Antennae by
computing the rms radial velocity dispersion of the 14 targets in the
overlap region.  We find a value of $74$ km s$^{-1}$ rms, which is
comparable with a disk galaxy's rms integrated velocity dispersion \citep{bottema93}.
%\citep[e.g. NGC~488,][]{gerssen97}.  
Some of this velocity dispersion is due to a general gradient in the north-south direction along the overlap region between the two nuclei \citep{amram92,gilbert02}, but the gradient is not monotonic: both nuclei have higher radial velocities of at least 1600 \kms\ \citep[][]{amram92}, and some ELCs lie at minima in \ha\ radial velocity, i.e. in \ha\ superbubbles \citep{amram92,whitmore99,zhang01}.  In the northern overlap region, the ELC rms radial velocity dispersion is 21 \kms\ (for eight ELCs) about a mean of 1573 \kms.  In the southern overlap region, the velocity dispersion of six younger ELCs is 48 \kms, equivalent to that of the giant molecular complexes, \citep[49 km
s$^{-1}$,][]{wilson00} about a mean of 1445 \kms, so the kinematics of this group reflect those of the nearby molecular material. 
Velocity dispersions for smaller groupings on scales of $10-20\arcsec$, e.g. the two B, three D, five E, or three F clusters, are 21, 56, 22, and 18 \kms, respectively, which apart from the larger D cluster value are comparable with the raw dispersions for sources over similar spatial scales in \citet{whitmore05}.

High-resolution near-IR spectroscopy has recently revealed broad supersonic \brg\ emission from massive embedded star-forming regions in two other systems:
\citet{turner03} observed a deeply embedded radio source in NGC~5253 and \citet{henry07} observed two similar sources in Henize $2-10$, for which they derived Q[H$^+$] values above $10^{52}$ s$^{-1}$ and FWHMs of order $60-80$ km s$^{-1}$.  
While these dwarf galaxies are different in scale from the merging Antennae disks, their starbursts have much in common:  optically visible SSCs that have already disrupted their ISM on large scales (seen e.g. in X-ray and \ha\ superbubbles), bELCs that appear to be blowing out of their natal cocoons, and even younger heavily enshrouded star-forming regions seen as radio-bright UDHIIs. 
% refs there? see hiir section below

%their SSCs have blown multiple X-ray superbubbles \citep{strickland99}, and harbors just a handful of SSCs \citep[e.g.][]{calzetti97} and one heavily enshrouded young star-forming region that may be a UDHII and a bELC. 

\section{Ages \& Masses of ELCs} 
\label{sec:agemass}

Br$\gamma$ fluxes trace the ionized gas within a cluster and hence
the young, massive stellar content.  We assume that this flux arises
from the O/WR stars of a population that formed in a short-lived burst
and has an initial mass function (IMF) that is usually characterized as a power law, $MdN/dM \propto M^{-\alpha}$.  Employing a stellar population
synthesis model such as Starburst99 \citep{leitherer99} permits an
inference of ELC age by comparing observed and predicted Br$\gamma$
equivalent widths (EWs) for a cluster as a function of age, and an
inference of mass follows from scaling the predicted magnitude of the
fiducial $10^6$ \Msun\ cluster at the determined age to match the
observed magnitude.  We assume solar metallicity for the
clusters\footnote{We note that while \citet{mengel02} found a
  handful of Antennae SSCs to be consistent with having solar
  metallicity, they argue that one (ELC S, their [W99]-2) may have a
  supersolar metallicity.} and adopt the Galactic field star average
IMF of Kroupa \citep[a broken power law with $\alpha = 2.3 $ for
stellar masses $M = 0.5 - 100$ \Msun\ and $\alpha = 1.3 $ for $M = 0.1
- 0.5$ \Msun,][]{kroupa01}.  The slope of this IMF at the high-mass
end is nearly equivalent to that of the solar neighborhood IMF of
Salpeter \citep[a single power law with $\alpha =
2.35$,][]{salpeter55}, and is consistent with that measured by
\citet{massey98a} for the massive, dense cluster R136 in the Large
Magellanic Cloud.  

Those assumptions yield ELC ages of $3.4-7$ Myr and masses of $7 \times
10^4 - 12 \times 10^6$ \Msun (incorporating an extinction correction
where possible), as listed in Table~\ref{tab:agemass}.  The ELC ages
have a gradient across the overlap region: the five ELCs that are
younger than 6 Myr reside in the southern overlap region where the
bulk of the molecular gas is concentrated \citep[e.g.][]{wilson00},
and the older sources reside in the northern overlap region.  We
estimate fit errors for the ages of less than 0.1 Myr and relative
errors in masses of 20\% for most ELCs (see Table~\ref{tab:agemass}).
However, more significant than the fit errors may be systematic
errors, possible sources of which include the unknown $K$-band
extinction for some ELCs, our assumptions that the ELCs are
ionization-bounded, and uncertainties in the models.  We discuss these
possibilities in turn below, and then compare our ages for two ELCs with literature values.

From an ELC \brg\ line flux we derive 
Q[H$^+$] under case B assumptions for an ionized \hii\ region at $10^4$ K
\citep[e.g.][]{osterbrock74,hummer87}.  The resulting values  
are listed in Table~\ref{tab:agemass} where they are uncorrected for extinction, 
whereas in the text we always adopt the listed extinction corrections where available.
ELC B1 has Q[H$^+$] $=1.5 \times
10^{53}$ s$^{-1}$, so it emits nearly half of the total Lyman
continuum rate from all 17 ELCs observed: $3.3 \times 10^{53}$
s$^{-1}$.  ELCs B1
and B have Q[H$^+$] at the high end of the range observed for SSCs in
other systems \citep[e.g. He $2-10$ region A at $10^{52-53}$
s$^{-1}$,][]{johnson00}.  UDHIIs are also observed to have Q[H$^+$]
values of order $10^{51-53}$ s$^{-1}$ \citep{beck00}.  The faintest
Antennae ELCs in our sample have Q[H$^+$] $= 10^{51}$ s$^{-1}$, which
is comparable with the fluxes of the brightest Galactic giant \hii\ 
regions (e.g. W43 and NGC~3606).

ELCs can suffer significant extinction even at $K$ band, and where estimates are
available we include them in Table~\ref{tab:agemass}.   In the case of ELC B1,
\citet{gilbert00} measured an extinction at $K$ band of ${A}_{\rm K} = 1.1 \pm
0.1 $ mag from Br$\gamma$ and 20 Pfund recombination lines.
This corresponds to ${A}_{\rm V} = 10$ mag for a standard extinction
law \citep{rieke85}.  Other authors have inferred lower extinctions to
this source: \citet{mengel01a} derived ${A}_{\rm V} = 4.3$ mag by
comparing \brg\ and \ha\ fluxes; \citet{whitmore02} report ${A}_{\rm
  V} = 7.62$ mag from optical imaging.  We use radio continuum flux
densities for ELC counterparts \citep[from][]{neff00} together with
\brg\ fluxes to estimate $K$-band extinctions for a few ELCs (B1, B, D2, and F1) under the
assumption that the radio emission is unabsorbed.  For ELC B1, our
inference of ${A}_{\rm V} = 10$ mag \citep{gilbert00} from K-band data agrees with
the extinction derived using the radio-IR flux ratio, suggesting that
% add detail about mixing
the emitting nebular gas and obscuring dust in the ELC are probably mixed; 
optical measurements only probe part-way into the emitting region so they underestimate the overall extinction and hence the intrinsic ${Q}
[{\rm H}^+]$.  \citet{turner03} find the same bias for the compact, heavily
obscured source 
in NGC 5253: near-IR and radio measurements give a consistent value
for the extinction of ${A}_{\rm V} = 18$ mag, a value that is six
times greater than the optically-determined one.  We emphasize that
infrared-radio determinations of ${Q} [{\rm H}^+]$ are more reliable
than optical-infrared ones.  In Tables~\ref{tab:measurements} and
\ref{tab:agemass} we present observed fluxes and derived masses (uncorrected
for extinction).  Extinction does not affect the ages since
they are derived from extinction-independent equivalent widths.

There is evidence that ionizing photons escape from young clusters in
starburst HIIRs \citep[e.g. in He $2-10$ and
NGC~4214,][]{johnson00,leitherer96a}.  Photoionization models of GHIIRs
require $10-73\%$ of the Lyman continuum flux to escape
\citep{castellanos02}.  
The escape fraction of Lyman continuum photons from Galactic OB associations is
estimated at $6-15\%$ \citep{dove00}.  Observations of diffuse
H$\alpha$ emission throughout the Antennae \citep[not only in
association with stellar clusters, e.g.][]{whitmore99} may also
suggest leakage of Lyman continuum flux from its HIIRs, i.e. that ELCs
are density-bounded.
\citet{beckman00} find that a transition between ionization-bounded
and density-bounded HIIRs in spiral galaxies occurs at a constant \ha\ 
luminosity of $L_{\ha} = 10^{38.6}$ erg s$^{-1}$.  For Case B
recombination this corresponds to a \brg\ luminosity $L_{\brg} =
10^{36.6}$ erg s$^{-1}$, which is below the observed values for all of
the ELCs. Presumably such a threshold in a merging pair of disk
galaxies is higher than in a spiral because it depends on the
available gas supply and the strength of the starburst, both of which
are stronger in the overlap region of a merger.
If ELCs are density-bounded, then some of their Lyman continuum photons
escape to ionize and dissociate the surrounding medium, and our
Br$\gamma$ measurements would underestimate $Q[H^+]$.
We would thus overestimate ELC ages and expect them to be fainter,
which would also lead to overestimated masses.

For two sources in our sample, \citet{mengel02} measured $K$-band
magnitudes and dynamical masses based on high-resolution spectroscopy,
and used additional diagnostics with EW[\brg], i.e. EW[CO] and
EW[CaT], to derive ages for the clusters.  We compare our ages with
theirs to give an idea of the significance of IMF choice, metallicity,
and the level of systematic errors that may be inherent to the observations and
Starburst99 models, which we both employ.
For ELC S (their cluster [W99]2), we infer an age of $6.6 \pm 0.1$ Myr
using EW[\brg], assuming a full ($0.1-100$ \Msun) Kroupa IMF at solar metallicity.
\citet{mengel02} obtain the same age ($6.6 \pm 0.3$ Myr) for a
Salpeter IMF ($1-100$ \Msun) at twice solar metallicity.  Our
measurements of EW[\brg] differ by a factor of two (ours: $4.8\pm1.6$
\AA, theirs: $10.0\pm2.5$ \AA), most likely because we use different apertures:
 theirs is a 2\farcs2 box that includes strong diffuse line emission surrounding ELC S
(which is not at a peak in \ha\ or \brg), while our aperture  encompasses
less  diffuse line emission because it is the best-fit Gaussian
spatial profile of the cluster \brg\ line, or 0\farcs9 FWHM for ELC S.  In
this case the differences in ages due to these other differences
(between IMFs, abundances, and even EW[\brg]) cancels out to be insignificant.
However, the typical size of the age offsets due to these factors is a few tenths of a Myr in
the sense that the Salpeter IMF gives an older age than the Kroupa IMF for a given EW[\brg], and a 
larger EW[\brg] for the same IMF gives a younger age, while the metallicity difference 
in age is quite small for this range of EW[\brg].
But \citet{mengel02} infer that neither of these IMFs may be appropriate for the
cluster based on the dynamical mass measurement ($2.0 \times 10^6$
\Msun), which is lower than the photometric mass predicted by the
models. This implies that the cluster has less than a full complement
of low-mass stars or a shallower IMF slope \citep{mengel02}.

In ELC F (their [W99]15), \citet{mengel02} measure a high EW[CO
2.29$\mu$m] of $17 \pm 0.2$ \AA.  This requires an age $>$ 7 Myr, when the
brightest supergiants dominate a cluster's infrared emission.  They
estimate an age for ELC F of $8.7 \pm 0.3$ Myr for a Salpeter IMF,
while we find $6.5 \pm 0.1$ Myr for a Kroupa IMF or ($6.6 \pm 0.1$ Myr 
for the same Salpeter IMF), where we both assume solar metallicity.
Again the dynamical mass implies that neither of these IMFs is
correct, but in the opposite sense to that of ELC S.  \citet{mengel02}
measure a dynamical mass of $3.3 \times 10^6$ \Msun, which suggests a
steeper than Salpeter IMF slope (2.5 rather than 2.35).  
The discrepancy between ages determined from \brg\ and those
determined from photospheric features of supergiants cannot be
explained by our different IMF choices or different magnitudes or
extinctions (these are the same).  It may be due to different values of EW[\brg] 
determined from different types of data: a spectrum in this work versus narrow-band
imaging in theirs, which are subject to different difficulties in
removing diffuse stellar and nebular backgrounds.  The discrepancy may
also be due to diverging model predictions for our differing spectral
diagnostics.  Strong \brg\ from O stars and CO absorption in
supergiants are not found simultaneously in the spectrum of a
single-aged stellar population based on the latest tracks used in Starburst99.
However, these are single-star evolutionary tracks, so the models do
not yet account for binary systems (in which most stars are thought to
reside) or the interactions between binaries and stars in 
cluster environments, e.g. mass transfer, wind-wind interactions, etc.
Finally, SSCs may not form instantaneously, but rather over the course of several Myr, and if SSCs observed at large distances are actually unresolved clusters of clusters then they could have larger age spreads--on scales similar to the ELC sizes, 30 Doradus has an age range of about 25 Myr due to propagating star formation: 
R136 contains low-mass stars with ages of $4-5$ Myr and massive stars with ages of $1-2$ Myr \citep{massey98}, while nearby Hodge 301 ($\sim 40$ pc away) harbors supergiants with ages of $20-25$ Myr \citep[e.g.][]{grebel00}.
We view ELC F and 30 Dor as a warning that ages determined via single-star, single-aged population synthesis models may have systematic errors of a few Myr at ages near
the end of the O star era, and proceed with this caveat in mind.

\section{Mass Loss from ELCs}
\label{sec:linewidths}
 
With mass estimates for the ELCs in hand, we now
assess the ability of the ELCs to gravitationally bind their \hii\
gas by comparing the observed \brg\ line widths $\sigma_{HII}$ of ELCs with their
escape velocities $v_{esc}$, which depend upon cluster mass and size.  We can write $v_{esc}$ in terms of the 1D stellar velocity dispersion $\sigma_*$ of a virialized cluster as $v_{esc}=\sqrt{2} \sigma_*$.  The virial relation can be expressed as $\sigma_*^2 = GM/\eta r_{hp} $, where $r_{hp}$ is the projected half-light radius and $\eta$ is a constant that depends upon the mass distribution of the cluster \citep[e.g.][]{spitzer87}.  The value of $\eta$ is often taken to be about 10, although for a range of realistic globular cluster concentration parameters, $0.5-2.5$, \citet{mengel02} found that $\eta=  9.7-5.6$ (more concentrated clusters have smaller $\eta$).  \citet{boily05} calculated the time evolution of $\eta$ in SSCs using a gas-dynamical model that incorporates stellar evolution; they found that in low-density clusters, $\eta$ does not change much over the first 10 Myr, but in intermediate- and high-density clusters that resemble M82-F and R136 in 30 Doradus, respectively, $\eta$ increases by factors of about $1.3$ and $1.7$, while the clusters grow more compact and $r_{hp}$ decreases.  An increasing $\eta$ implies a decreasing $v_{esc}$, which makes it progressively easier for \hii\ gas to escape from a cluster.  
If the measured $\sigma_{HII}$ equals or exceeds $v_{esc}$, then the cluster cannot gravitationally bind the high-velocity ionized gas in the line wings, which has sufficient energy to escape from the cluster potential in an unbound outflow (in the absence of a confining external pressure). 
For the two ELCs (S and F) whose $\sigma_*$ were measured by \citet{mengel02},
we can directly compare $v_{esc}$ and $\sigma_{HII}$.  Assuming that the measured $\sigma_*$ represents a virial velocity dispersion, ELC S has $v_{esc}=20.1$ 
%$\sigma_* = 14.2$
\kms\ and $\sigma_{HII} = 31.8$ \kms\ (corrected for the instrumental line profile, Table~\ref{tab:measurements}), so most of its gas is not bound to the cluster.  If the other ELCs in
Table~\ref{tab:measurements} share this typical value for an SSC
stellar velocity dispersion, then nearly all of them have
predominantly unbound gas.  ELC F has the second-lowest intrinsic nebular line width at
$\sigma_{HII} = 11.6$ \kms, and it has $v_{esc} = 28.6$ \kms,
%$\sigma_*=20.2$
which suggests that most of its gas is bound to the cluster.  If mass segregation is present in these ELCs, then the measured $\sigma_*$ would be an underestimate of the true value because it is dominated by the light of massive, centrally concentrated stars.  In that case, ELCs S and F would be better able to bind their nebular gas.

\begin{figure}[th] 
\begin{center}
%\epsscale{0.75}
\plotone{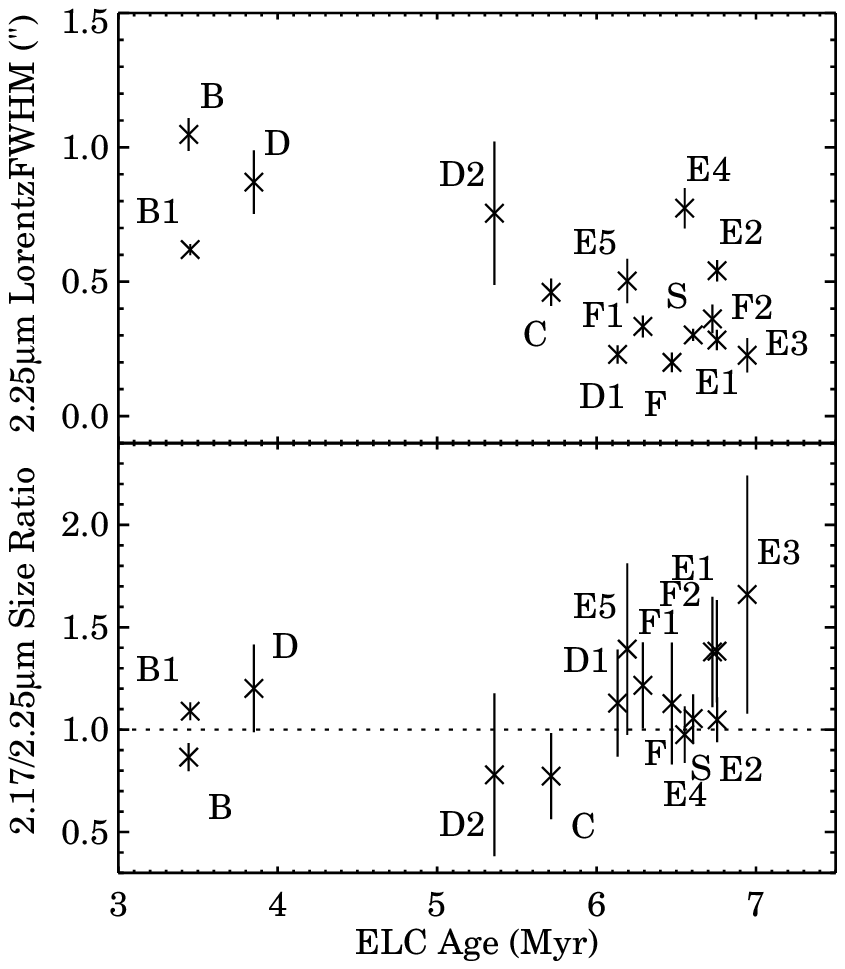}
\caption[Sizes of Antennae ELC as function of age] {Above: Deconvolved sizes of ELCs measured from 2.25 \um\ ISAAC narrow-band image, plotted as a function of ELC age.   At ages of $6-7$ Myr, the population of ELCs has a large size spread (down to half-light radii of 9 pc for ELC F), but younger ELCs are larger, with radii of $30-50$ pc.  Below:  The ratio of 2.17 \um\ (stellar continuum plus \brg\ emission) to 2.25 \um\ (stellar continuum) half-light radii for bELCs increases slightly with age.  Most bELCs (which exclude ELCs F and D1) have larger sizes in the \brg\ filter than in the continuum one.  Sizes are geometric mean FWHMs determined from 2D Lorentzian fits with Star 4 geometric mean radius subtracted in quadrature, and error bars are computed from fit errors. \label{fig:sizes}}
\end{center}
\end{figure}

In order to estimate escape velocities for the other ELCs, we
consider the mass of both the cluster's stellar component (implied by
Starburst99 for a Kroupa IMF extending from 0.1 to 100 \Msun,
\S~\ref{sec:agemass}) and its ionized gas component (whose mass ranges from 0.01 to 2.6 times the stellar mass\footnote{The ionized gas mass is deduced from outflow model fits in Paper II, \S 2.3 and Table 1.}).  We adopt an extinction correction where possible
(as in Table~\ref{tab:agemass}), and we assume that stars and gas fill the same volume out to the smaller of the half-light radii measured from the 2.17 and 2.25 \um\
FWHMs (see Figure~\ref{fig:sizes} and Table~\ref{tab:phottable}).  These assumptions are conservative in the sense that they place the total amount of mass available in the minimum volume, which leads to an upper limit for $v_{esc}$.   
Adopting a value of $\eta=10$, we derive escape velocities for most sources that are well below the observed \hii\ line widths: $v_{esc}/\sigma_{HII}$ ranges from 0.1 to 0.3 for all ELCs except the two clusters with the narrowest \brg\ emission (i.e., all bELCs have $\sigma_{HII} > v_{esc}$).  
The exceptions are ELC F, which we have already concluded can bind its gas, and
ELC D1, which has the narrowest \brg\ line in the sample, and $v_{esc}/\sigma_{HII} = 2.1$.  
A more conservative estimate of the gas boundedness comes from assuming that most of the ELC mass is actually concentrated in a 4 pc radius \citep[median value for SSC effective radii in][]{whitmore99} that is unresolved in our data; this leads to higher values of $v_{esc}/\sigma_{HII} = 0.2-0.8$  for bELCs, but still suggests that much of the HII gas in the observed profiles is not bound to the clusters.
If we consider both this smaller radius and decrease $\eta$ to the minimum value of 5.6 that corresponds to a high-concentration globular \citep{mengel02}, then the ratios $v_{esc}/\sigma_{HII}$ for ELCs B1, B, and E5 rise to $ 0.9- 1.1$, while those of the remaining bELCs are in the range $0.2- 0.8$.  Thus even with the most compact mass distribution that we imagine they could have, even in the ELCs with the largest $v_{esc}/\sigma_{HII}$ ratios, the high-velocity gas in the broad line wings would be unbound.
This suggests that a significant fraction of the ionized gas in bELCs is unbound and may be outflowing, but that narrow-lined ELCs can gravitationally bind their HIIRs and do not drive outflows.  

An exception to this rule may be the compact, embedded broad-line source in NGC~5253 \citep{turner03} that resembles the Antennae bELCs in terms of line width, mass, and luminosity.  Although its \brg\ line is very broad (76 \kms\ FWHM) and hence supersonic, it appears to be compact enough to bind its \hii\ gas if the most conservative choices for $\eta$ and half-light radius (the smallest radio semimajor axis is 0.5 pc) are adopted for a $10^6$ \Msun\ cluster \citep{turner03}; in this case, $v_{esc}=131$ \kms, which greatly exceeds the measured line width, and leads to the conclusion that the cluster easily binds its UDHII.  However, if we assume $\eta = 10$ and use the $K$-band size of the nebula to delineate the \brg-emitting region \citep[1 pc radius for their H2, ][]{alonsoherrero04} rather than the radio size, then we find $v_{esc}=69$ \kms.  This would suggest that plenty of gas in the \brg\ line wings has adequate energy to escape the cluster's potential well.  Either way, the NGC~5253 cluster is far more compact than the Antennae ELCs; this difference may be due to a higher ISM pressure in NGC~5253 that confines the UDHII at high density and prevents any significant outflow of gas.
% mention he 2-10 sources here??

\begin{figure}[th] 
\begin{center}
\epsscale{0.75}
\plotone{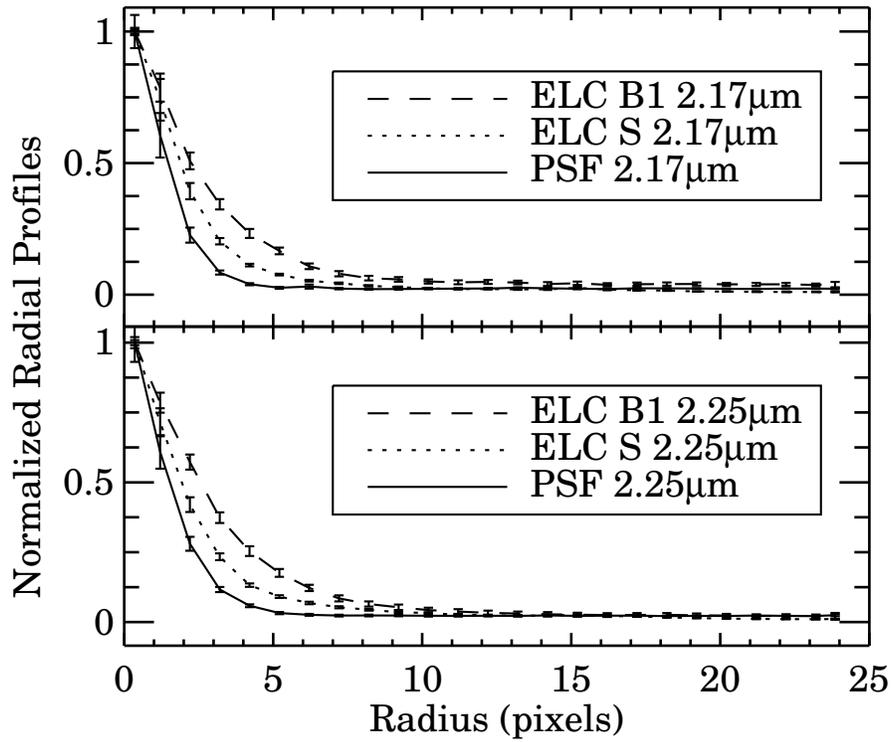}
\caption[Antennae ELC light profiles]{Normalized, azimuthally averaged radial profiles of ISAAC 2.17 and 2.25 \um\ image counts for two ELCs and Star 4 demonstrate that the ELCs are resolved relative to the observed PSF.  Error bars indicate $1\sigma$ Poisson errors.
\label{fig:radprofs}}
\end{center}
\end{figure}

% radii shift slightly
In the foregoing discussion we have assumed that the stars in ELCs are distributed like their infrared light.  The $K$-band half-light radii of ELCs range from about 10 to 50 pc, measured either in the continuum narrow-band filter at 2.25 \um, which is dominated by starlight (Table~\ref{tab:phottable}), or in the 2.17 \um\ filter, which also includes \brg\ nebular emission (Figure~\ref{fig:sizes}).  ELC radii are up to an order of magnitude larger than the median effective radius of optically visible SSCs of all ages in the Antennae (4 pc), measured from HST images \citep[][and see Figure~\ref{fig:ant_hst}]{whitmore99}. \citet{mengel02} measured an effective radius of $3.6$ pc for the
apparent optical counterpart of ELC F from the $I$-band HST
observations, and they found similar values up to 6 pc for a few other
optically revealed clusters.  However, \citet{mengel05} reported that the mean $I$-band effective radius is 16 pc (with a standard deviation of 15 pc) for seven isolated young (age $< 4$ Myr) Antennae clusters.  The 2.25 \um\ half-light radii of ELCs B1, B, and D, whose ages are below 4 Myr, are 29, 49, and 40 pc, respectively (Table~\ref{tab:phottable}, Figure~\ref{fig:sizes}).  Thus young clusters are intrinsically large at visible wavelengths even at the WFPC2 resolution of $0\farcs1$.  Most ELCs are resolved in our near-infrared images (with point source FWHM of $0\farcs4$ in the combined frames, see Fig.~\ref{fig:radprofs}), but what we view as a single source could break up into multiple sources when viewed with better angular resolution.  
In order to check this, we examined archival optical and near-IR HST images of several ELCs together with ISAAC ones: Figure~\ref{fig:thumbs} shows that while some ELCs consist of multiple sources, even at the finest spatial resolution they tend to be dominated by a single bright one.
ELC B, for example, is comprised of a very bright source in $K$ band surrounded by fainter emission and at least one neighboring source to the southwest, but there are a handful of spatially coincident sources in HST filter F555W \citep[Figure~\ref{fig:ant_hst} inset,][]{whitmore99}; it may be a cluster of clusters.  ELC B1 is a single bright source (like most other ELCs) in $K$ band and a very faint one in $I$.  Caution must be exercised in attempting to match sources observed at different wavelengths and resolutions, especially when extinction can be important.
% add ref to hst fig 2, fig:thumbs, 

\begin{figure}[th] 
\begin{center}
\epsscale{0.75}
\plotone{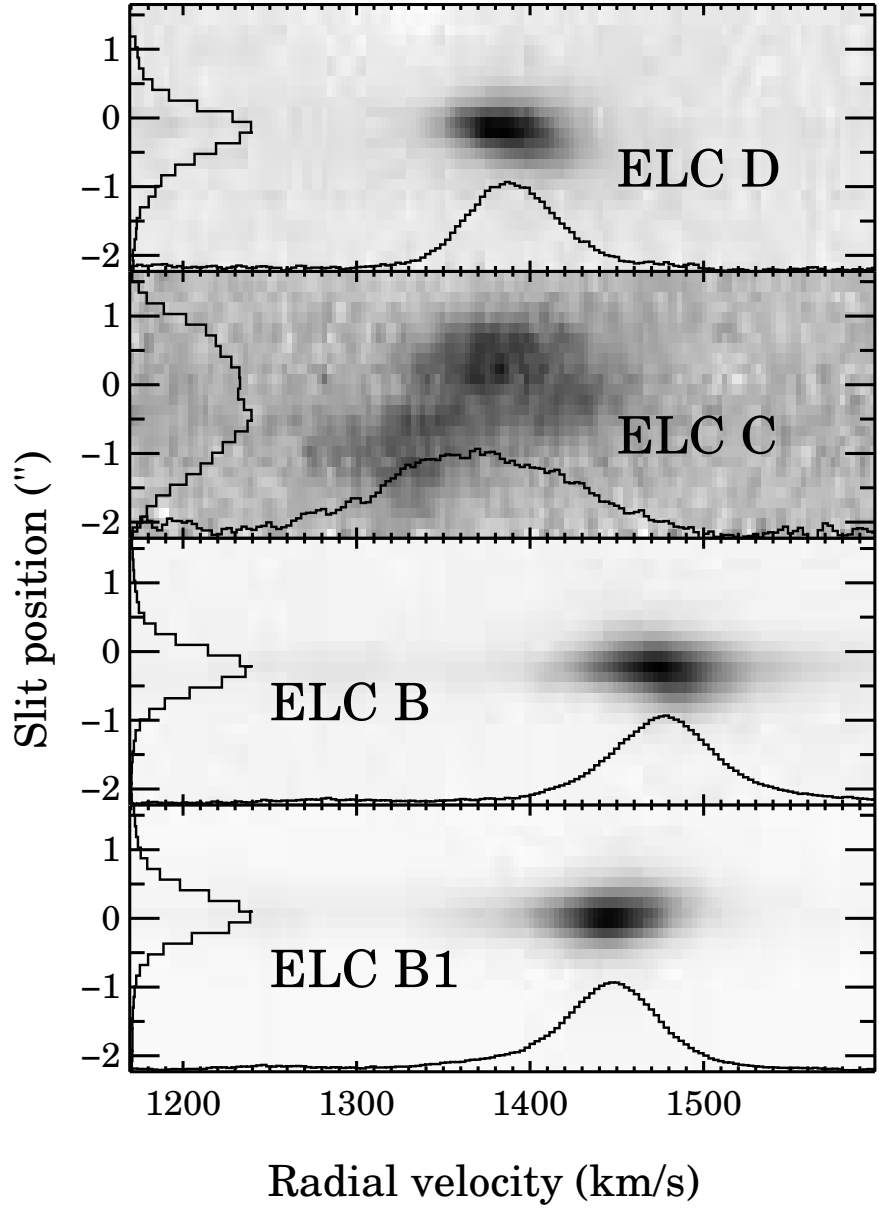}
\caption[Br$\gamma$ position-velocity diagrams for Antennae ELCs] {Br$\gamma$
  position-velocity diagrams for ELCs B1, B, C, and D, overlaid with
  arbitrarily scaled integrated spatial profiles and extracted spectra.  
  The \brg\ emission is 
  extended with respect to the continuum, and its spatially resolved
  velocity gradients are suggestive of nonspherical flows.
\label{fig:elcpv}}
\end{center}
\end{figure}

The position-velocity diagrams in Figure~\ref{fig:elcpv} show that the
\brg\ emission in ELCs B1, B, and D has spatial and velocity structure.  ELC C, which
appears complex and diffuse in optical and near-IR images, also has
extremely extended spatial and velocity structure in \brg\, with a
very weak continuum that is centered between its high- and low-velocity 'lobes'; it may comprise several cluster components, or it may be driving a bipolar outflow.  Figure~\ref{fig:sizes} shows that most (but not all) bELCs have slightly more extended emission in the 2.17 \um\ filter, which includes \brg, than in the 2.25 \um\ filter, and that there is a weak tendency for the ratio of 2.17 \um\ to 2.25 \um\ sizes to increase with age.  An HIIR might have more extended line than continuum emission (which
is dominated by stars) because Lyman continuum photons are escaping to
ionize the surrounding ISM, and this may produce some of the observed
extended emission of ELCs.  However, another process that would
produce extended line emission is the escape of ionized gas in an
outflow, which is suggested by the high velocities of some of the
extended \brg\ relative to the line centers, as well as the double-lobed appearance of ELC C in Figure~\ref{fig:elcpv}.  An increase in \brg-to-continuum size ratio with bELC age could be due to expansion of an outflow into the surrounding ISM and/or contraction of the stellar cluster with time (see \S~\ref{sec:evolution}).

\section{The Nature of ELC HIIRs}
\label{sec:densities}
\label{sec:lumsig}

It is useful to compare the HIIRs that are excited by ELCs with three familiar catagories of HIIRs 
(Galactic compact and ultra-compact HIIRs ((U)CHIIs), Galactic HIIRs, and extragalactic GHIIRs), as well as with the recently discovered class of UDHIIs \citep{kobulnicky99}.  
We will discuss four properties that distinguish different types of HIIRs: luminosity (in the form of $Q[H^+]$), line width, size, and electron density $n_e$.

\begin{figure}[th]
\begin{center}
\epsscale{1.15}
\plottwo{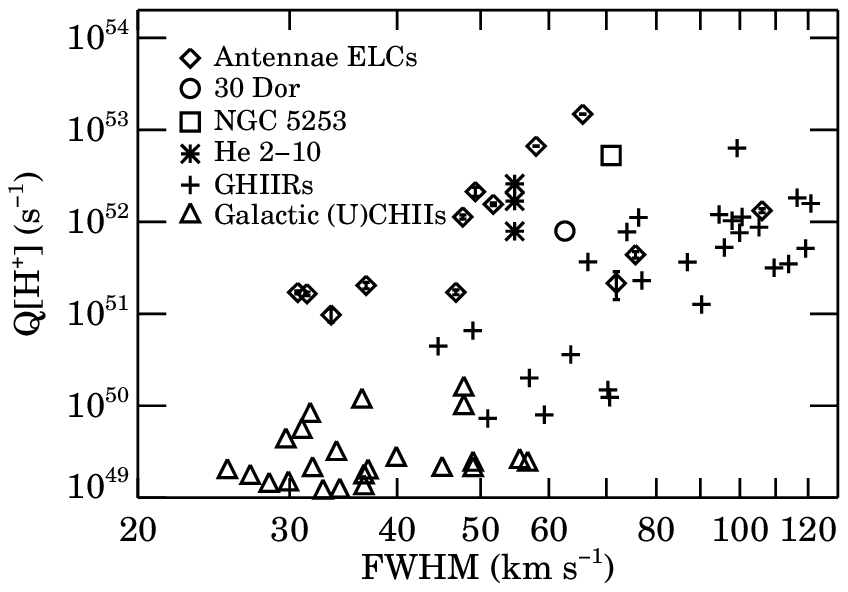}{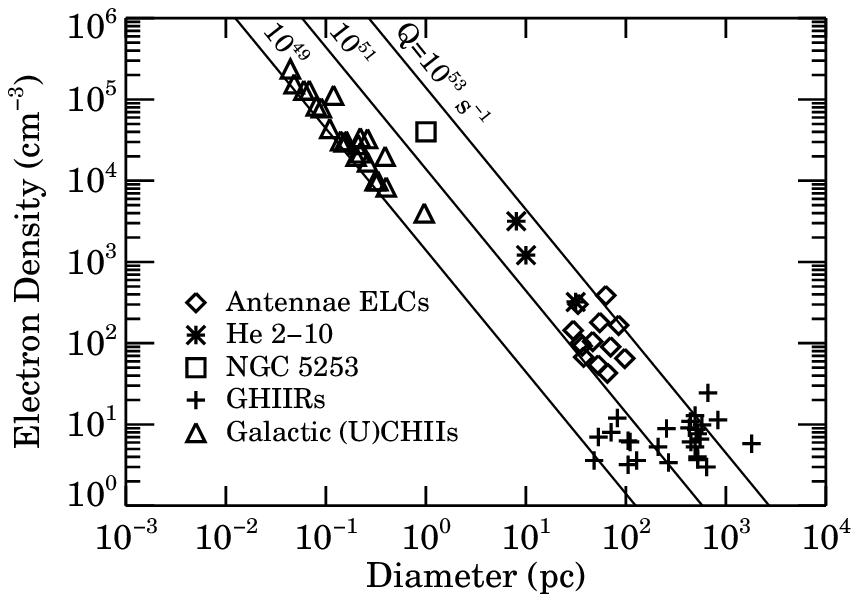}
\caption[Relation between bELC Lyman continuum rate and linewidth, and electron density and diameter] {Above: bELC Lyman
  continuum rate versus line width, with instrumental response and thermal component removed, compared with other samples of HIIRs:
  nearby massive cluster 30 Dor \protect{\citep[luminosity and line width
  from][respectively]{walborn91,chu94}, UDHIIs/bELCs in NGC 5253
  \citep{turner03} and He $2-10$ \protect{\citep{kobulnicky99,johnson03,vacca02,henry07}}, Galactic (U)CHIIs
  \citep[][]{garay99}, and first-ranked GHIIRs in several galaxies
  \citep[][]{arsenault88}.   Below: Mean electron density versus diameter for bELCs compared with the other HIIR populations.  Solid lines indicate the Str\"omgren sphere relation for an HIIR with constant density: $n_{e} \propto {d}^{-1.5}$, for $Q[H^+] = 10^{49}, 10^{51}$, and $10^{53}$ s$^{-1}$}.  Sizes are measured from radio data for all but the bELCs and UDHIIs, whose sizes are from near- and mid-IR images, respectively; adopting the pc-scale radio sizes for UDHIIs or the typical optical SSC sizes for bELCs moves them closer to the (U)CHIIs.  All densities plotted are average values; the GHIIRs would move up by one or two orders of magnitude if we adopted the higher densities that are derived from optical forbidden lines and are weighted toward dense, presumably low-filling-factor gas.}
\label{fig:lumsig}
\end{center}
\end{figure}

We presented $Q[H^+]$ values for ELCs in \S~\ref{sec:agemass} and discussed
the \brg\ line widths of ELCs in  \S~\ref{sec:linewidths}; we now place bELCs (excluding the narrow-lined  ELCs D1 and F) in context with other HIIRs by plotting their $Q[H^+]$ vs.
 H~{\sc i} recombination line widths (with instrumental and thermal components, assuming a $10^4$ K HIIR, removed),  together with data for other objects in Figure~\ref{fig:lumsig}. The luminosity of 30 Doradus in the LMC is taken from \citet{walborn91} and its integrated \ha\ line width
inferred from the spatially resolved echelle spectroscopy of
\citet{chu94}. The luminosity and \brg\ width for the ELC-like
source in NGC~5253 are from \citet{turner03}.  
For the UDHIIs in He $2-10$ the $Q[H^+]$ values are derived from the radio data of \citet{johnson03}, and the line width adopted is the typical value of about 55 \kms\ observed for \brg\ (corrected for instrumental and thermal contributions, which resembles bELC line widths) in several slits that overlap with several of the radio sources \citep{henry07}.
% added He 2-10 widths
The line widths for
a sample of (U)CHIIs  from \citet{garay99} are for H~{\sc i} radio recombination lines.
The sample of top-ranked  GHIIRs in nearby galaxies is from \citet{arsenault88}, who compiled \ha\ line widths, luminosities, rms $n_e$ values, and diameters from the literature.  We assume case B recombination to determine $Q[H^+]$ values from line fluxes.
Figure~\ref{fig:lumsig}  shows bELCs at the high end of 
the range in $Q[H^+]$ and FWHM, where they overlap with the GHIIRs, which are characterized by supersonic linewidths \citep{smith70} and $Q[H^+] \sim 10^{50}-10^{52}$ s$^{-1}$.  
%
% UDHIIs are not shown in the $Q[H^+]$-FWHM plot because line widths are not available for them, but their $Q[H^+]$ values are very high, $\sim 10^{52}$ s$^{-1}$ \citep[for the UDHIIs in He $2-10$,][]{kobulnicky99}, which would place them among the bELCs and brightest GHIIRs.
%
Galactic (U)CHIIs have much lower luminosities because they are powered by just one or a few massive stars \citep[\protect{$Q[H^+] \sim 10^{44}-10^{50}$ s$^{-1}$},][]{garay99}, but their line widths overlap with those of the brighter HIIRs. 
Not shown are common Galactic HIIRs, which exhibit narrower H recombination line FWHMs of $20-40$ km s$^{-1}$ and  $Q[H^+] = 4 \times 10^{47} -
10^{51}$ s$^{-1}$ \citep{churchwell78}.  
 A loose correlation between FWHM and $Q[H^+]$ is evident, and its slope for bELCs is consistent with that reported for GHIIR populations of nearby spirals \citep[i.e. NGC~4449 and NGC~4321,][]{arsenault90,rozas98,fuentesmasip00a}.

We also compare measurements of size (diameter $d$) and $n_e$ for several types of HIIRs in Figure~\ref{fig:lumsig}.  The variations in $d$ and $n_e$ within the (U)CHII, UDHII, and bELC populations extend along tracks of constant $Q[H^+]$ for a constant-density Str\"omgren sphere: $n_e \propto d^{-1.5}$ (solid lines in Figure~\ref{fig:lumsig}), while GHIIRs cover a broader range in $Q[H^+]$ values.  While GHIIRs are similar to bELCs in terms of luminosity or inferred stellar mass, they are  larger ($d \gs 100 $ pc) and much less dense \citep[e.g. ELC B1 has \protect{$n_e = 10^{3.5}- 10^4$ cm$^{-3}$ from [Fe~{\sc iii}]} ratios,][and see below]{gilbert00}.  GHIIRs have rms densities of a few to a few $100$ cm$^{-3}$ (derived from emission measure profiles), but forbidden-line diagnostics, e.g. [S~{\sc ii}] and [O~{\sc ii}], yield densities that are ten times greater \citep[e.g.,][]{kennicutt84}.  This range of densities is commonly  interpreted as evidence for clumpiness of the HIIRs, and it suggests filling factors of dense gas of $0.01-0.1$ \citep{kennicutt84}.  Such a variable-density medium in GHIIRs is certainly plausible: inhomogeneities are common in resolved HIIRs like 30 Dor \citep[e.g.]{chu94}, and density variations have been measured in both Galactic and extragalactic HIIRs  \citep{copetti00,castaneda92}.

(U)CHIIs are included in Figure~\ref{fig:lumsig} because they represent extremes in all of the plotted parameters.  Although their low $Q[H^+]$ values, small implied stellar masses, and sub-parsec sizes clearly differentiate Galactic (U)CHIIs from GHIIRs and bELCs, their extremely high densities are similar to that of ELC B1:  compact (diameter $d < 0.5$ pc) and ultra-compact ($d < 0.05$ pc) HIIRs have $n_e$ ranging over $\sim 10^3 - 10^5$ cm$^{-3}$ \citep{wood89}.
Common Galactic HIIRs (not plotted) have comparable luminosities to those of (U)CHIIs, but they are larger ($0.5-15$ pc) and have slightly lower densities ($10^2 - 10^4$ cm$^{-3}$) than their compact cousins.
Moreover, UCHIIs have dynamical ages indicating extreme youth \citep[5000 years,][]{wood89}, although they could live much longer ($\sim 10^5$ years) \citep{garay99}, so the UCHIIs may be the precursors of common HIIRs.  The bELCs appear to be $3-7$ Myr old (\S~\ref{sec:agemass}) and therefore much more evolved than UCHIIs, yet they are clearly more compact and perhaps at an earlier stage of evolution than the diffuse GHIIRs.  

We estimate a volume-averaged $n_e$ for all ELCs by assuming ionization and recombination balance for a $10^4$ K HIIR with Q[$H^+$] (extinction-corrected, where possible) from Table~\ref{tab:agemass} along with their \brg\ sizes (i.e. the FWHM measured in our 2.17 \um\ narrow-band image, see Figure~\ref{fig:sizes}).
 The resulting values range from 30 to 600 cm$^{-3}$ with an average of 180
cm$^{-3}$, and they represent lower limits on the electron densities
of ELCs.  For the four bELCs that have thermal radio counterparts, \citet{neff00} compute similar rms $n_e$ values of $30-360$ cm$^{-3}$ by assuming measured diameters of up to $70-120$ pc (which in some cases gives an upper limit on size and lower limit on $n_e$ because the sources are not resolved by their $\sim 80$ pc beam). 
\citet{neff00} also consider the possibility that their thermal radio sources have $8-30$ pc diameters that are nearer to those observed for SSCs \citep[$\sim 8$ pc,][]{whitmore99}, and thereby obtain higher estimates of $n_e = 1 - 5 \times 10^3$ cm$^{-3}$.  If the $K$-band emission from ELCs is concentrated in compact, unresolved cores that
are the size of the optical SSCs, then the density estimates would increase correspondingly, to $700-8500$ cm$^{-3}$ with a mean of 2250 cm$^{-3}$; the maximum value is that of ELC B1, and it is consistent with the range inferred from [Fe~{\sc iii}] ratios by \citet{gilbert00}. 
Thus the radio-inferred densities for a subset of ELCs are comparable with the IR-inferred ones.  For consistency, the $n_e$ values plotted in Figure~\ref{fig:lumsig} are the lowest (rms $n_e$) estimates for GHIIRs, ELCs, and UDHIIs, which we discuss next.

We estimate densities for the UDHIIs of He
$2-10$ in the same way as for the ELCs, using the $Q[H^+]$ values from 0.7 cm data of
\citet{johnson03} and the 10\um\ sizes of \citet{vacca02}, where we
combine their objects 1 and 2 since they are blended in the observations of the latter work.
Thus we assume that the mid-IR and near-IR sizes are comparable, and that the stars and warm dust share the same volume.
We find densities for the two compact UDHIIs (4 and 5, with radii
of 4 and 5 pc) of 2200 and 1280 cm$^{-3}$.  For the two larger UDHII
regions (1+2 and 3, with radii of 16 and 17 pc), we find $n_e \approx
390 $ and 180 cm$^{-3}$.  The lowest estimates of average ELC densities are thus similar to those
of the more extended UDHIIs in He $2-10$, and the SSC-size estimates are similar to those of the compact UDHIIs.  

However, \citet{vacca02} do not assume that the stars and warm dust in UDHIIs share the same volume; they estimate $n_e$ for the He $2-10$ UDHIIs by fitting the radio-NIR SEDs. \citet{vacca02} model a UDHII
as a central HIIR that envelopes a young SSC and extends out to the
inner radius of a surrounding annular dust cocoon that extends to an
outer radius.  Their best-fit inner radii are similar to the sizes of
optical SSCs, $2-5$ pc, and they are consistent with fits to more recent radio data of \citet{johnson03}.  This leads to average values for $n_e$ of
$1400-5720$ cm$^{-3}$, which agree well with the radio estimates of
the average $n_e$ from \citet{kobulnicky99} and \citet{johnson03}.
For the compact radio counterpart of the NGC~5253 HIIR,
\citet{turner00} report an even larger estimate of the rms $n_e$ of $
4 \times 10^4$ cm$^{-3}$, which is in the range of UDHII $n_e$.
Adopting the higher estimates of $n_e$ for UDHIIs or bELCs\footnote{We discuss further evidence for high-electron-density material in bELCs in
Paper II, where we derive radial density distributions for the bELC
HIIRs by fitting the \brg\ profiles with a kinematic model for a
cluster outflow; this yields even higher values for the
emissivity-weighted average $n_e$: $10^3-10^5 $ cm$^{-3}$.}
 would move them toward to the (U)CHIIs in the $n_e-d$ panel of Figure~\ref{fig:lumsig}, while adopting the higher estimates for GHIIRs would move them up; the figure would still display a strong anticorrelation between $n_e$ and diameter in which bELCs and UDHIIs occupy a central region that is distinct from that of the other massive systems plotted, the GHIIRs.

% add a caveat about sample diffs/selection effects
Although the populations of HIIRs that are discussed above exist in a variety of environments and were all observed in different ways and subject to different selection effects from ELCs, their comparison shows how ELCs fit into the broader context of star formation.
Their nearest relatives are the UDHIIs, whose Lyman
continuum luminosities and electron densities are similar, but UDHIIs are so heavily obscured that they are rarely visible in the optical range and are thought to be younger than ELCs.  UDHIIs are classified based on their radio properties, and ELCs based on their recombination lines, but we have already conflated the two groups by assuming the \brg\ line fluxes of \citet{henry07} emerge from the radio/MIR sources of \citet{kobulnicky99} and \citet{vacca02}.  That is because all evidence suggests that ELCs and UDHIIs represent the same objects, perhaps at slightly different evolutionary stages, so that massive clusters form within gas- and dust-shrouded UDHIIs, which become ELCs as their massive hot stars drive winds (that are visible in recombination line emission) and expel gas to reveal young SSCs with low extinctions at visible wavelengths.  If these massive clusters form hierarchically from clusters of subclusters \citep[e.g.][]{testi00,bonnell03}, then the ELCs may be comprised of several evolving UDHIIs that our data cannot resolve.
We conclude that bELCs and UDHIIs have some properties in common with
other types of HIIRs, but taken together, their broad
lines, high densities, intermediate sizes, and large luminosities constitute a new class
of massive HIIRs that is distinct from GHIIRs.  
% "objects" changed to "massive hiirs"

% added par:
While our small sample spans two orders of magnitude in $Q[H^+]$ and inferred cluster mass, it may represent only the bright end of the full young cluster population in the Antennae.  If the mass function of young clusters in the Antennae starburst reaches down to single-star scales, would those objects resemble Galactic UCHIIs?  The analogy between UCHIIs and UDHIIs may extend beyond extreme densities, compactness, and extinctions:  the massive stars within UCHIIs are thought to spend $10-20$ \%  of their main-sequence lifetimes in an embedded phase \citep{wood89}, and those in UDHIIs may spend even longer fractions of their lives hidden from optical view, e.g. 15 \% in He~$2-10$ \citep[][]{kobulnicky99} to 40 \% in NGC~5253 \citep[][]{martinhernandez05}, until their cluster winds remove much of the obscuring gas.  Such an extended embedded phase for massive stars in starbursts is proposed  to explain discrepancies between measured mid-IR line ratios and stellar population synthesis models with normal upper mass limits by \citet{rigby04}, who point out that the higher ISM pressures in starbursts should lengthen their embedded compact HIIR phase relative to that in the Galaxy.  The implications of a long embedded phase in which IR nebular lines are suppressed are serious for starburst population synthesis modeling \citep{rigby04}.
% and the interpretation of the correlation between UV slope and IR-to-UV flux ratio in starbursts \citep{vacca02}.

\section{Evolution of ELCs}
\label{sec:evolution}

% mean sizes shift a bit
In an evolutionary sequence determined by cluster age and embeddedness, massive stellar clusters form in heavily obscured UDHIIs, evacuate their surroundings as wind-blowing ELCs, and, if they survive the mass-loss phase, emerge as SSCs, which may ultimately become GCs if they survive further destructive cluster dynamical processes and tidal interactions with their host galaxy.  This scenario requires some explanation of cluster size evolution, in particular how the large $K$-band radii of ELCs ($10-50$ pc, Table~\ref{tab:phottable} and Figure~\ref{fig:sizes}) can be reconciled with the typical half-light radii of optically revealed SSCs \citep[$\sim 4$ pc in the Antennae,][]{whitmore99}.  We assume that ELCs are young SSCs embedded in HIIRs because their inferred stellar masses are similar to those of SSCs, which can be factors of a few to ten times more compact than ELCs.  Figure~\ref{fig:sizes} shows a wide scatter in the 2.25 \um\ stellar continuum sizes of 10 ELCs at ages of $6-7$ Myr: their mean ($\pm$ standard deviation) half-light radius is $17 \pm 8$ pc.  Younger ELCs (5 have ages of $3-6$ Myr) have a larger mean half-light radius of $35 \pm 11$ pc at 2.25 \um.   Additional evidence for a decrease in Antennae cluster size with age, as well as a large scatter in size, is presented by \citet{mengel05}, who report average ($\pm$ standard deviation) WFPC2 $I$-band effective radii for 18 isolated clusters with ages below 4 Myr and in the range $8-11$ Myr of $16 \pm 15$ pc and $6.5 \pm 5.3$ pc, respectively.  

This tendency of older clusters to be more compact could be explained by cluster contraction with age or by the preferential survival of more concentrated clusters at any given age.  Clusters that form hierarchically would contract as subclusters merge to form a single cluster.  Simulations by \citet{bonnell03} show that star clusters may form hierarchically, such that fragmentation in a molecular cloud produces sheets and filaments which form stars before the rest of the cloud material, and these stars form high-density subclusters that merge after several free-fall times to create a centrally condensed cluster.  Observations of the Serpens molecular cloud core provided the first evidence for hierarchical cluster formation: \citet{testi00} find that star formation is not homogeneous in the core, but rather is concentrated in dense subclusters that are well separated spatially and kinematically.  If ELCs are young enough to be undergoing the final merger of their initial subclusters, this might explain their size trend in Figure~\ref{fig:sizes}.  

However, even if hierarchical formation is still underway in ELCs and causing contraction, it must compete with the countervailing influence of mass loss due to stellar winds that lead to cluster outflows.  Cluster expansion is caused by the loss of gas via stellar evolution and stars via relaxation, tidal stripping, and dynamical friction.  Mass loss and consequent expansion can unbind a cluster and cause it to dissolve below detection limits into the background stellar population \citep[see][for an analysis of these effects across the cluster mass spectrum]{fall01}; hence compact clusters live longer than less compact ones, and they increasingly dominate the cluster population as they age.  Two recent studies of the age distribution of stellar clusters in the Antennae argue that the sharp decrease in the number of clusters as a function of age is strong evidence for rapid dissolution of the majority of young clusters on timescales of order 10 Myr \citep{mengel05,fall05}.  Thus we favor expansion and rapid dissolution over contraction due to hierarchical formation in explaining the smaller sizes of older ELCs.

Large ELCs may therefore be puffed-up young SSCs that have expanded due to prodigious mass loss, which limits their chances for survival if they are not compact enough, if their IMF is top-heavy or truncated at low masses, or if their star-formation efficiency (SFE) is not high enough.  \citet{boily03b} simulate cluster evolution in the presence of rapid (i.e. instantaneous) mass loss; after equilibrium is re-established, they find expansion factors for the half-mass radii of up to a factor of ten or more for bound clusters having Plummer and King distribution functions with varied concentrations and assumed SFEs.  This is a plausible explanation for the sizes of ELCs if they were born as more compact clusters within UDHIIs; however, the mass loss in ELCs proceeds over the course of at least several Myr, so they would not suffer such extreme expansion as the clusters simulated by \citet{boily03b}. 

The expansion of young clusters due to dynamical evolution is observed in nearby resolved clusters: cluster core radii in the LMC and SMC increase with age, at varying rates, up to ages of at least 1 Gyr \citep{elson89,mackey03b,mackey03a}.  These young clusters ($10^6-10^9$ yrs old) have power-law luminosity profiles that often show bumps or dips that suggest the presence of substructures that have not been smoothed out by relaxation \citep{elson87,mackey03a,mackey03b}.  These profiles do not display the tidal truncation of a King profile \citep{king66}, which best describes the luminosity profiles of much more evolved GCs.  There is evidence that the more massive SSCs in nearby galaxies also have power-law light profiles, some extending beyond 100 pc in radius, some with clumpy substructure \citep[see references in][]{schweizer04}; evidence for the onset of tidal truncation in older SSCs (beyond a few 100 Myr) and intermediate-age GCs has also been found \citep{whitmore99, schweizer04}.  \citet{larsen04c} finds that extended envelopes relative to King profiles are common among the youngest clusters in a sample of spiral galaxies observed with HST WFPC2, but his sample does not show the trend of increasing cluster radii with age that is seen in the LMC and SMC. 

The interpretation of large ELCs as puffed-up SSCs that are caught in the act of dissolving assumes that their $K$-band light at large radii is dominated by stellar continuum emission.  Alternative explanations for their sizes might, however, involve dust scattering and emission from a shell at larger radii surrounding an SSC-sized cluster, 
or differential extinction that preferentially obscures the cluster core relative to its outer radii, making the observed FWHM larger than the intrinsic one.  A final possibility is that SSCs may form hierarchically from clusters of smaller subclusters that merge rapidly to make a single massive cluster \citep[within a couple of free-fall times for a low-mass cluster, ][]{bonnell03};  in this picture, prior to the final merger into a single cluster, the average stellar density  is smaller than in individual subclusters, and the overall structure is probably not spherically symmetric.  Observations at higher angular resolution at infrared wavelengths, either space-based or adaptive optics-assisted, will be able to test these scenarios.

\section{Star Formation in Massive Clusters}
\label{sec:sfr}

Comparing the star-formation rate (SFR) in ELCs with that of the entire system is difficult
because the SFRs of galaxies are determined from observed quantities
such as \ha\ and far-infrared (FIR) luminosities averaged over large
spatial scales, and they consider continuous star formation
% and a Salpeter IMF down to 0.1 \Msun\ 
\citep[e.g.][]{kennicutt98}.  Assuming the calibrations of
\citet{kennicutt98}, adopting an IRAS luminosity of $10^{11}$ \Lsun\ 
for the Antennae leads to an SFR$_{\rm FIR} \approx 17$ \Msun\ 
yr$^{-1}$; \citet{zhang01} use Kennicutt's \ha\ calibration to
estimate SFR$_{\ha} \approx 20$ \Msun~yr$^{-1}$.  If we apply this
\ha\ calibration to the sample of ELCs by assuming case B
recombination, we derive a total SFR$_{\ha}$ of 3.2 \Msun\ yr$^{-1}$.
Thus the fraction of the recent SFR of the Antennae
that is provided by ELCs is at least 15\%, and half of this
contribution is provided by ELC B1 alone.  This is a lower limit because we have neglected extinction in two-thirds of the ELCs, our sample probably does not include all ELCs, and there could be a significant escape fraction for ionizing photons from the clusters.
From a photometric study of the current star cluster population in the Antennae, \citet{mengel05} estimate a total SFR in clusters over the past 25 Myr of at least 20 \Msun\ yr$^{-1}$ (in agreement with the \ha\ estimate for all star formation), but they argue that the age distribution of SSCs implies rapid dissolution of most clusters, and that accounting for this could raise the SFR to at least 50 \Msun\ yr$^{-1}$.

An estimate of the fraction of stars that are formed in clusters rather than in the field may be derived from imaging observations that compare amounts of discrete and diffuse emission from
star-formation tracers: \citet{fischer96} reported that about 70\% of
the \brg\ emission in a 30\arcsec\ aperture including the nucleus of
NGC~4039 and the southern overlap region is diffuse, while the
remaining 30\% is associated with \ha\ and radio knots.  Similarly,
\citet{hummel86} report that the total fraction of radio continuum
emission from discrete knots (at 6\arcsec\ resolution) is about 35\%,
with errors of $20-30$\%.  Our lower limit to the fraction of the
star-formation rate in ELCs is about half of what one might expect
from these imaging studies.  These results are consistent with the range deduced from UV observations of samples of starbursts: about 15--50\% of the UV light
is typically associated with stellar clusters
\citep[e.g.][]{meurer95b,maoz96}.  Radio continuum imaging of the
Wolf-Rayet starburst He 2-10 reveals that 15\% of the O stars in the
galaxy reside in UDHIIs \citep{kobulnicky99}, which are probably the
progenitors of ELCs and SSCs. Since our measurement is a lower limit
on the fraction of star formation in Antennae clusters, it appears
that ELCs constitute the majority of the {\it clustered} star-formation rate
in the Antennae, and at least 15\% of the \ha-derived total SFR.

The actual proportion of clustered star formation in the Antennae may be 
nearer to 100\% if many of the young SSCs are quickly dissolved or disrupted \citep[e.g.][]{mengel05,fall05}.  \citet{whitmore99} find a power-law form for the luminosity function for Antennae SSCs that closely resembles that of other SSCs in other starbursts \citep[see e.g.][]{meurer95a,larsen02} as well as that of OB associations and GHIIRs \citep[e.g.][]{kennicutt89,mckee97}.  Many of the low-mass SSCs must dissolve into the field population in order for a power-law SSC population to evolve into a log-normal globular cluster population; simulations indicate that this is plausible \citep{fall01}, and observations have begun to reveal the process \citep[e.g.][]{degrijs03,lamers05}.
Indeed, \citet{mengel05} and \citet{fall05} argue that rapid dissolution into the field population is the normal fate for SSCs in the Antennae; those that survive mass loss in the ELC phase are subject to further destructive dynamical and evolutionary processes that may prevent them from becoming GCs.  

\section{Conclusions}
\label{sec:conclusions}
High-resolution spectroscopy of the youngest SSCs in the Antennae
reveals broad, spatially extended, non-Gaussian, supersonic \brg\ 
emission ($60-110$ km s$^{-1}$ FWHM) in the ELCs.  Typical GHIIRs also have supersonic linewidths and
luminosities at the low end of the ELC ranges, but they are much
larger and less dense than ELCs, and are generally excited by
lower-mass clusters and OB associations.  Thus ELCs appear to
represent a different class of dense extragalactic HIIR that is powered by young SSCs and may be associated with high-pressure environments.  Their closest relatives
are the UDHIIs, which are younger (age $\ls 1$ Myr), more heavily
embedded radio and mid-IR nebulae with high electron densities. UDHII Lyman continuum luminosities and infrared sizes are similar to those of ELCs, so UDHIIs
require similar stellar masses for their embedded ionizing clusters as ELCs,
$10^5-10^7$ \Msun.  Thus UDHIIs are the probable precursors of ELC.  

The youngest ELCs (ages $3-6$ Myr) are concentrated in the southern
overlap region of the Antennae, while those in the northern overlap
region are slightly older ($6-7$ Myr).  The rms radial velocity
dispersion of the youngest ELCs is consistent with
that measured for molecular gas via CO in the same region, so the
young clusters and dense gas share similar kinematics.  ELCs in our incomplete sample comprise about 15\% of the current SFR in the Antennae (half of that is concentrated in the brightest cluster) and they make up a significant fraction of the expected clustered star formation in the system.  

Because bELC \brg\ linewidths typically exceed the most conservative estimates of the cluster escape velocities, much of the H~{\sc ii} gas is not bound to the
clusters--it is flowing out in a cluster wind.  The greater extent of \brg\ emission relative to continuum emission is consistent with escape of ionized gas in winds.
Although this mass loss takes place over the course of several Myr rather than instantaneously, it may be dramatic enough to prevent some of the ELCs from surviving as bound SSCs; their large $K$-band sizes and the decrease in mean ELC size with age may be the signature of rapid cluster expansion and dissolution, which leads to the preferential survival of the most compact clusters in each age group.  In Paper II we present a kinematic model for cluster outflows to infer their mass-loss rates and energetics, and discuss ELC feedback in the context of similar systems such as wind-blown superbubbles and galactic
superwinds.

%%%%%%%%%%%%%%%%%%%
% Acknowledgments %
%%%%%%%%%%%%%%%%%%%

\acknowledgments

We thank M. Lehnert and W. Vacca for discussions and comments, and we thank the 
referee for comments that improved the paper. 
We also  thank the staffs of the European Southern Observatory and the
Keck Observatory, and
observing assistant Ron Quick in particular.  The authors wish to
recognize and acknowledge the very significant cultural role and
reverence that the summit of Mauna Kea has always had within the
indigenous Hawaiian community.  We are most fortunate to have the
opportunity to conduct observations from this mountain.  

This work was supported in part by National Science Foundation (NSF)
grant AST-0205999, a NASA GSRP fellowship, and the NSF Science and
Technology Center for Adaptive Optics, managed by the University of
California at Santa Cruz under cooperative agreement No. AST-9876783.
Part of this work was performed under the auspices of the U.S. Department of Energy,
National Nuclear Security Administration by the University of California,
Lawrence Livermore National Laboratory under contract No. W-7405-Eng-48.

%%%%%%%%%%%%%%
% References %
%%%%%%%%%%%%%%

%\bibliographystyle{apj} 
%\bibliography{ssc}

% Fig 1 revised new names, inset size correction, mag x5 not x8

% add a new Fig 2 in revised version, name change

% Fig 2-> 3 revised names, order

% Fig 3->4 revised names

% Fig. 5 added in revised version

% was Fig 4, becomes 6 
% lorentz_sizes.eps -> f6.eps
% figure/numbers updated with plate scale corrections.  new fits, errors, names. C1 -> d1.
% 7 -> 9 pc.  25-40 pc -> 30-50 pc.

% Fig 5a,b--becomes 7a,b, revised sizes--add note that width has instr/thermal components removed, also specify that bELCs not all ELCs are plotted.
  
\clearpage

% TAb 1--modified
% revised 4-07 with new sizes (lorentz deconvd), names, star4 fwhm (lorentz geommean)
\begin{deluxetable}{lrrrrr} 
\tablewidth{0pt} 
\tablecaption{Photometry \& Sizes of Selected Antennae Sources
\label{tab:phottable}} 
\tablehead{ 
 \colhead{SSC}  & \colhead{$\Delta$ R.A.\tablenotemark{a}} & \colhead{$\Delta$ Dec\tablenotemark{a}} & \colhead{K\tablenotemark{b}} & \colhead{FWHM\tablenotemark{c}} \\
Name & ('') & ('') & (mag) & ('')\\
}
\startdata
   B1 & -17.7 & -21.0 &  14.7 & 0.62 \\
    B & -23.1 & -18.2 &  14.5 & 1.05 \\
    D & -11.7 &  -4.5 &  15.4 & 0.87 \\
    C & -20.2 &  -6.3 &  15.2 & 0.46 \\
   D1 & -13.9 &  -2.7 &  15.8 & 0.23 \\
   D2 & -10.3 &  -1.1 &  16.9 & 0.75 \\
   E1 & -11.7 &  15.5 &  16.1 & 0.28 \\
   E2 & -13.0 &  14.7 &  15.6 & 0.54 \\
   E3 & -19.5 &  11.4 &  17.1 & 0.23 \\
    F & -11.9 &  26.6 &  15.9 & 0.20 \\
   E5 &  -8.9 &  21.8 &  15.3 & 0.50 \\
   F2 & -15.9 &  32.5 &  15.8 & 0.36 \\
   F1 & -13.9 &  28.8 &  15.7 & 0.33 \\
   E4 & -10.7 &  20.4 &  15.7 & 0.77 \\
   A1 & -39.8 & -23.0 &  17.5 & 0.32\tablenotemark{d} \\
    S & -82.3 &  32.1 &  14.0 & 0.30 \\
\enddata
\tiny
\tablenotetext{a}
{Positions in arcseconds relative to Star 4
of Whitmore et al. 1999, whose position is reported as R.A. = $12^h
1^m 56.04^s$ and DEC = $-18^\circ 52' 43\farcs66$ (J2000), although
\citet{whitmore02} find an offset of \protect{$1\farcs2$} to the
southwest of the HST positions from radio observations
\cite{neff00}.  All offsets are measured from our NIRSPEC N7 mosaic 
 except that of S, which is derived 
from \citet{whitmore99}.}
\tablenotetext{b}
{K magnitude from NIRSPEC SCAM photometry, except for SSC S
from \citet{mengel02}.  Systematic errors are 0.2 mag. }
\tablenotetext{c}
{Deconvolved FWHM (from geometric mean radius of 2D Lorentzian fit) from
  distortion-corrected 2.25 \um\ narrow-band ISAAC image;
 at the Antennae \protect{$1\arcsec = 93$} pc. 
 Lorentzian FWHM of Star 4 is 0.27\arcsec, which 
 was used to derive seeing-deconvolved FWHMs.}
\tablenotetext{d}
{A1 is too faint and confused to fit as above, so its FWHM is estimated from a spline fit to the azimuthally averaged profile, subtracting in quadrature Star 4's size measured the same way (0.38\arcsec).}
\end{deluxetable}

\clearpage

% Table 2
% names altered 4-07.  what about cross-refs for other literature names?
\begin{deluxetable}{lcr}
\tablewidth{0pt}
\tablecaption{Log of NIRSPEC Spectra of Antennae Sources \label{tab:obselc}}
\tablehead{
 \colhead{Source} &  \colhead{Date}   & \colhead{$t_{exp}$}  \\
\multicolumn{1}{c}{Names} &  (UT)  & \multicolumn{1}{c}{(s)}\\
}
\startdata
 B,B1\tablenotemark{a}  & 2002-02-22  & 2 $\times$ 600 \\
 C,D  & 2002-02-22  & 2 $\times$ 600 \\
 D1,D2   & 2002-02-22 & 1 $\times$ 600 \\
 E1,E2,E3   & 2002-02-22 & 3 $\times$ 600 \\
%  10,16    & 2002-02-22 & 1 $\times$ 600 \\
 F\tablenotemark{b}    & 2002-02-23 &  2 $\times$ 600 \\
W99 10,16\tablenotemark{c}  & 2002-02-23  & 2 $\times$ 600 \\
 F1,F2,E5   &  2002-02-23   & 3 $\times$ 600 \\
 E4   &  2002-02-23  & 1 $\times$ 600 \\
N4039 nucleus, A1 &  2002-02-23  & 1 $\times$ 300 \\
N4038 nucleus &  2002-02-23 & 1 $\times$ 300 \\
 S\tablenotemark{d} & 2000-12-11 & 7 $\times$ 900 \\
 W99 1\tablenotemark{c} & 2001-02-04 & 10 $\times$ 900 \\
\enddata
\tablenotetext{a}
{SSC B1 is SSC A of \citet{gilbert00}.}
\tablenotetext{b}
{Young cluster 15 of \citet{whitmore99}.}
\tablenotetext{c}
{Young clusters 10, 16, and 1 of \citet{whitmore99}.}
\tablenotetext{d}
{Young cluster 2 of \citet{whitmore99}.}
\end{deluxetable}

\clearpage

% Tab 3--modified
% names modified to match tab1. 4-07  reorder?
\begin{deluxetable}{lrrrr}
%\tabletypesize{\scriptsize}
\tablecolumns{5}
\tablewidth{0pt}
\tablecaption{ \brg\ Line Measurements
\label{tab:measurements}}
\tablehead{
\colhead{Name} & \colhead{F[Br$\gamma$]$\times 10^{16}$}\tablenotemark{a}  &\colhead{ EW[Br$\gamma$]} & \colhead{$v_{rad}$\tablenotemark{b}} &\colhead{ $v_{FWHM}$\tablenotemark{c}} \\
\multicolumn{1}{l}{} & {(erg s$^{-1}$ cm$^{-2}$)} &  \multicolumn{1}{c}{~~~(\AA)} & \multicolumn{2}{c}{(km s$^{-1}$)}
}
\startdata
   B1 &  155.7 $\pm$    2.4 & 254.8 $\pm$   8.0 &   1476 &    70\\
    B &  166.1 $\pm$    1.6 & 260.9 $\pm$   5.3 &   1447 &    63\\
    D &   47.5 $\pm$    1.7 & 189.7 $\pm$  10.6 &   1390 &    57\\
    C &   40.5 $\pm$    2.1 &  63.8 $\pm$   3.6 &   1383 &   109\\
   D1 &    3.1 $\pm$    0.6 &  24.2 $\pm$  10.4 &   1485 &    14\\
   D2\tablenotemark{d} &   30.9 $\pm$    7.2 & 123.2 $\pm$  52.4 &   1487 &    55\\
   E1 &    5.1 $\pm$    0.2 &   2.6 $\pm$   0.1 &   1599 &    40\\
   E2 &    5.2 $\pm$    0.2 &   2.5 $\pm$   0.1 &   1580 &    39\\
   E3 &    5.2 $\pm$    0.3 &   1.7 $\pm$   0.1 &   1544 &    53\\
    F &    6.7 $\pm$    0.5 &   8.4 $\pm$   0.7 &   1583 &    30\\
   E5 &    6.2 $\pm$    0.5 &  20.8 $\pm$   3.6 &   1558 &    44\\
   F2 &   13.4 $\pm$    1.1 &   2.8 $\pm$   0.3 &   1601 &    80\\
   F1 &    2.3 $\pm$    0.5 &  15.9 $\pm$   3.5 &   1566 &    42\\
   E4 &   34.5 $\pm$    1.8 &   6.0 $\pm$   0.4 &   1558 &    54\\
   A1\tablenotemark{e}  &   64.2 $\pm$    1.8 &   1.6 $\pm$   0.0 &   1575 &    45\\
N4038 &   33.4 $\pm$    3.5 &   1.6 $\pm$   0.2 &   1597 &    77\\
    S &    6.6 $\pm$    2.2 &   4.8 $\pm$   1.6 &   1566 &    76\\
\enddata
\tiny
\tablenotetext{a}{\protect{\brg} flux is not corrected for extinction 
(see Table~\ref{tab:agemass}).}
\tablenotetext{b}{Average error in barycentric \brg\ radial velocity 
is 1 \kms, with a maximum of 2.8 \kms.}
\tablenotetext{c}{Observed FWHM, not corrected for instrumental line-spread
  function (FWHM 12 \kms). This correction
is significant for the marginally resolved line of SSC D1.  
The corrected FWHM of SSC F is 27 \kms.
Average error in \brg\ FWHM is 2.6 \kms, 
with a maximum of 7.5 \kms.}
\tablenotetext{d}{For D2, total flux and EW are listed, 
but v$_{\rm rad}$ and v$_{\rm FWHM}$ refer only to the broad 
component of the line.}
\tablenotetext{e}{Refers to an off-nuclear HIIR near the NGC~4039 nucleus.}

\end{deluxetable}

\clearpage

% Tab 4--modified
%names modified to match tab1 4-07
\begin{deluxetable}{lrrrr}
\tablewidth{0pt}
\tablecaption{Derived Ages \& Masses  \label{tab:agemass}}
\tablecolumns{5}
\tablehead{
\colhead{Name} & \colhead{Log Q[H$^+$]\tablenotemark{a}} & \colhead{Age\tablenotemark{b,c}} & \colhead{Mass\tablenotemark{b,d}} & \colhead{$A_K$} \\
\multicolumn{1}{l}{} & \multicolumn{1}{r}{(s$^{-1}$)} & (Myr) & \multicolumn{1}{c}{(10$^6$ \Msun)}& (mag) 
}
\startdata
% Kroupa agemass values, .1-100 Msun, NOT INCLUDING extinction correction
% for Q and mass.  See sb99_agemass.pro
   B1 & 52.71 &  3.45 &  4.23 &  1.2\tablenotemark{e} \\
    B & 52.74 &  3.44 &  5.00 &  0.2\tablenotemark{e} \\
    D & 52.19 &  3.85 &  1.93 &  ...\\
    C & 52.12 &  5.72 &  4.13 &  ...\\
   D1 & 51.01 &  6.13 &  1.62 &  ...\\
   D2 & 52.00 &  5.36 &  0.80 &  0.8\tablenotemark{e} \\
   E1 & 51.22 &  6.75 &  0.26 & ...\\
   E2 & 51.23 &  6.76 &  0.41 & ...\\
   E3 & 51.23 &  6.94 &  0.07 & ...\\
    F & 51.34 &  6.47 &  0.74 &  0.1\tablenotemark{f} \\
   E5 & 51.31 &  6.19 &  2.60 &  ...\\
   F2 & 51.64 &  6.73 &  0.35 &  ...\\
   F1 & 50.87 &  6.29 &  1.47 &  0.3\tablenotemark{e} \\
   E4 & 52.05 &  6.55 &  0.65 &  ...\\
   A1 & 52.32 &  7.00 &  0.50 & \\
N4038 & 52.04 &  7.00 &  3.45 &  ...\\
    S & 51.33 &  6.61 &  3.22 & 0.0\tablenotemark{f} \\
\enddata
\tiny
\tablenotetext{a}{Q[H$^+$] is derived from \brg\ fluxes and
 not corrected for extinction, which scales it up by
 factor $10^{0.4{\rm A_K}}$.}
\tablenotetext{b}{Ages and masses (uncorrected for extinction)
 were derived by comparing
measured K magnitudes and \brg\ EWs with
Starburst99 models for a cluster with solar metallicity and Kroupa
IMF ($0.1-100$ \Msun). A nonzero A$_{\rm K}$ does not affect ages, but does
scale masses by factor $10^{0.4{\rm A_K}}$.}
\tablenotetext{c}{Errors in ages are derived from Starburst99 fits and 
are typically below 0.1 Myr.  For ELCs D2 and F1 the large errors on EW 
lead to larger age errors: D2 has age $5.4_{-1.4}^{+0.4}$ Myr and F1 has 
age $6.4_{-0.06}^{+1.8}$ Myr.}
\tablenotetext{d}{ Relative errors in masses are typically $20 \%$ and 
dominated by photometric errors for most sources.  Exceptions are ELCs 
D1, D2, and F1, whose relative errors reach 37, 30, and 80\%, 
respectively, due to their large age uncertainties. }
\tablenotetext{e}{ Estimated from \brg\ fluxes and radio measurements reported
by \citet{neff00} (see \S~\ref{sec:agemass}).}
\tablenotetext{f}{ Reported by \citet{mengel02}.}
\end{deluxetable}

\end{document}